\documentclass[10pt,preprint]{aastex}

\usepackage{mathtools}
\usepackage{amsmath}
\usepackage{relsize}
\usepackage{booktabs}
\usepackage{comment}
\usepackage{natbib}
\DeclarePairedDelimiter{\abs}{\lvert}{\rvert}

\usepackage{color}

\newcommand{\GHz}{{\rm GHz}}

\shorttitle{Ripples from the Recombination Epoch}
\shortauthors{Sathyanaraynara Rao et al.}

\begin{document}


\title{On the detection of spectral ripples from the Recombination Epoch}


\author{Mayuri Sathyanarayana Rao$^{1,2}$, Ravi Subrahmanyan$^{2}$, N Udaya Shankar$^{2}$, Jens Chluba$^{3}$}
\affil{{\small $^{1}$Australian National University, Research School for Astronomy \& Astrophysics, Mount Stromlo Observatory, Cotter Road, Weston, ACT 2611, Australia}}
\affil{{\small $^{2}$Raman Research Institute, C V Raman Avenue, Sadashivanagar, Bangalore 560080, India}}
\affil{{\small $^{3}$Department of Physics and Astronomy, Johns Hopkins University, \\3400 N. Charles Street, Baltimore, MD 21218, USA}}
\email{Email of corresponding author: mayuris@rri.res.in}

\begin{abstract}
Photons emitted during the epochs of Hydrogen ($500 \lesssim  z \lesssim  1600$) and Helium recombination ($1600 \lesssim  z \lesssim 3500$ for HeII $\rightarrow$ HeI, $5000 \lesssim z \lesssim  8000$ for HeIII $\rightarrow$ HeII) are predicted to appear as broad, weak spectral distortions of the Cosmic Microwave Background. We present a feasibility study for a ground-based experimental detection of these recombination lines, which would provide an observational constraint on the thermal ionization history of the Universe, uniquely probing astrophysical cosmology beyond the last scattering surface. We find that an octave band in the 2--6 GHz window is optimal for such an experiment, both maximizing signal-to-noise ratio and including sufficient line spectral structure. At these frequencies the predicted signal appears as an additive quasi-sinusoidal component with amplitude about $8$~nK that is embedded in a sky spectrum some nine orders of magnitude brighter. We discuss an algorithm to detect these tiny spectral fluctuations in the sky spectrum by foreground modeling. We introduce a \textit{Maximally Smooth} function capable of describing the foreground spectrum and distinguishing the signal of interest. With Bayesian statistical tests and mock data we estimate that a detection of the predicted distortions is possible with 90\% confidence by observing for 255 days with an array of 128 radiometers using cryogenically cooled state-of-the-art receivers. We conclude that detection is in principle feasible in realistic observing times; we propose APSERa---Array of Precision Spectrometers for the Epoch of Recombination---a dedicated radio telescope to detect these recombination lines.
\end{abstract}

\keywords{Astronomical instrumentation, methods and techniques - Methods: observational - Cosmic background radiation - Cosmology: observations  - Recombination - Radio continuum: ISM }

\section{Introduction}

The cosmological recombination epoch marks an extended period over which electrons recombine with protons and Helium nuclei as cosmological expansion and cooling cause the Universe to transition from a fully ionized primordial plasma to a gas of almost completely neutral Hydrogen and Helium atoms. In the cosmological concordance model, the cosmological recombination epoch spans redshifts $500 \lesssim z \lesssim 1600$ for Hydrogen and for Helium recombination the corresponding redshifts are $1600 \lesssim z \lesssim 3500$ for HeII $\rightarrow$ HeI and $5000\lesssim z \lesssim 8000$ for HeIII $\rightarrow$ HeII. At these redshifts, electrons and photons are tightly coupled through energy exchange via Compton scattering and the thermodynamic temperature of electrons and photons are almost exactly equal (at $T\approx 3815\{(1+z)/1400\}$~K). During Hydrogen recombination, once even a small neutral fraction builds up, photons emitted in any free-bound transition to the ground state would almost immediately be re-absorbed by a nearby neutral hydrogen atom, thereby effectively compensating for the electron captured \citep{Zeldovich68, Peebles68, Chluba2007b}.  Hydrogen recombination therefore relies on free-bound transitions to excited states followed by a trickle down to the ground state with the subsequent emission of numerous recombination photons.  During this process, atoms are frequently photo-ionized by the huge number of CMB photons and after multiple dissociations and recaptures the atoms finally reach the ground state, emitting photons in the Lyman-$\alpha$ resonance or the $2s$--$1s$ two-photon continuum \citep{Zeldovich68, Peebles68} and, at low frequencies, in transitions among excited states \citep{Dubrovich1975, Rybicki93, Dubrovich1997}.  

Recombination is expected to be stalled as the ambient radiation temperature in the Lyman-$\alpha$ transition wavelength rises and is balanced in a quasi-static equilibrium with populations in the 1$s$ and 2$p$ states.  Recombination thus depends on the depletion rate of $n=2$ atoms and removal of the excess brightness in the ambient Lyman-$\alpha$ line by Hubble expansion and two photon decay from the 2$s$ state. These processes occur at a rate that is $\simeq 10^7-10^8$ times slower than the spontaneous transition probability of the Lyman-$\alpha$ and play a vital role in controlling the dynamics of recombination. 
About $57\%$ of all hydrogen atoms become neutral through the $2s$--$1s$ two-photon channel \citep{Chluba2006b}, which has a vacuum decay rate of only $A_{2s1s}\simeq 8.22\,{\rm s^{-1}}$ \citep{Goeppert1931, Breit1940, Spitzer1951, Goldman1989, Labzowsky2005}. Consequently, hydrogen recombination is expected to be substantially delayed compared to what might be expected assuming equilibrium Saha recombination at the average densities and temperatures typical for our Universe.

 As the number of photons per baryon is roughly $1.6 \times 10^9$, radiative processes including stimulated recombination, induced emission and absorption of photons dominate the populating of atomic levels rather than collisions \citep[e.g.,][]{Chluba2007, Chluba2010}. As the Universe gradually expands and recombination proceeds by uncompensated bound-bound and free-bound transitions, the level populations of atomic species slowly fall out of equilibrium with the radiation field, which causes departures of the CMB spectrum from an ideal Planck form. The spectral lines corresponding to these transitions are predicted to appear as redshifted additive deviations to the cosmic microwave background (CMB) spectrum, with most of the Hydrogen lines originating from redshift $z\simeq 1300-1400$ \citep{Jose2006,Chluba2006b}; the observable intensity of these spectral lines is furthermore expected to be uniform on the sky and unpolarized. The lines are also substantially broadened owing to the extended recombination time (and electron scattering in the case of HeII recombination). At low frequencies, adjacent lines overlap substantially and hence the cosmological recombination radiation is expected to manifest itself as spectral `ripples' riding on a smooth continuum, which are together an additive component of the extragalactic background light extending over radio, microwaves and near IR wavelengths.

Improvements in our understanding of the recombination history within the framework provided by the concordance cosmology, the role of $2s$--$1s$ two-photon decay and other factors governing the depletion of the $n=2$ level population,  the importance of the contribution of Helium to the recombination process, progress in related atomic physics and, last but not the least important, substantial improvements in computing power have all contributed to much improved and more realistic modeling of the epoch in question \citep[see][for overview of recombination physics]{Fendt2009, Jose2010}.  Detailed calculations as described in \cite{Sunyaev2009} and references therein provide a fairly precise estimate of the spectral ripples expected due to the recombination of Hydrogen and Helium.   

\begin{figure}[h]
\centering
\includegraphics[width=15.0cm]{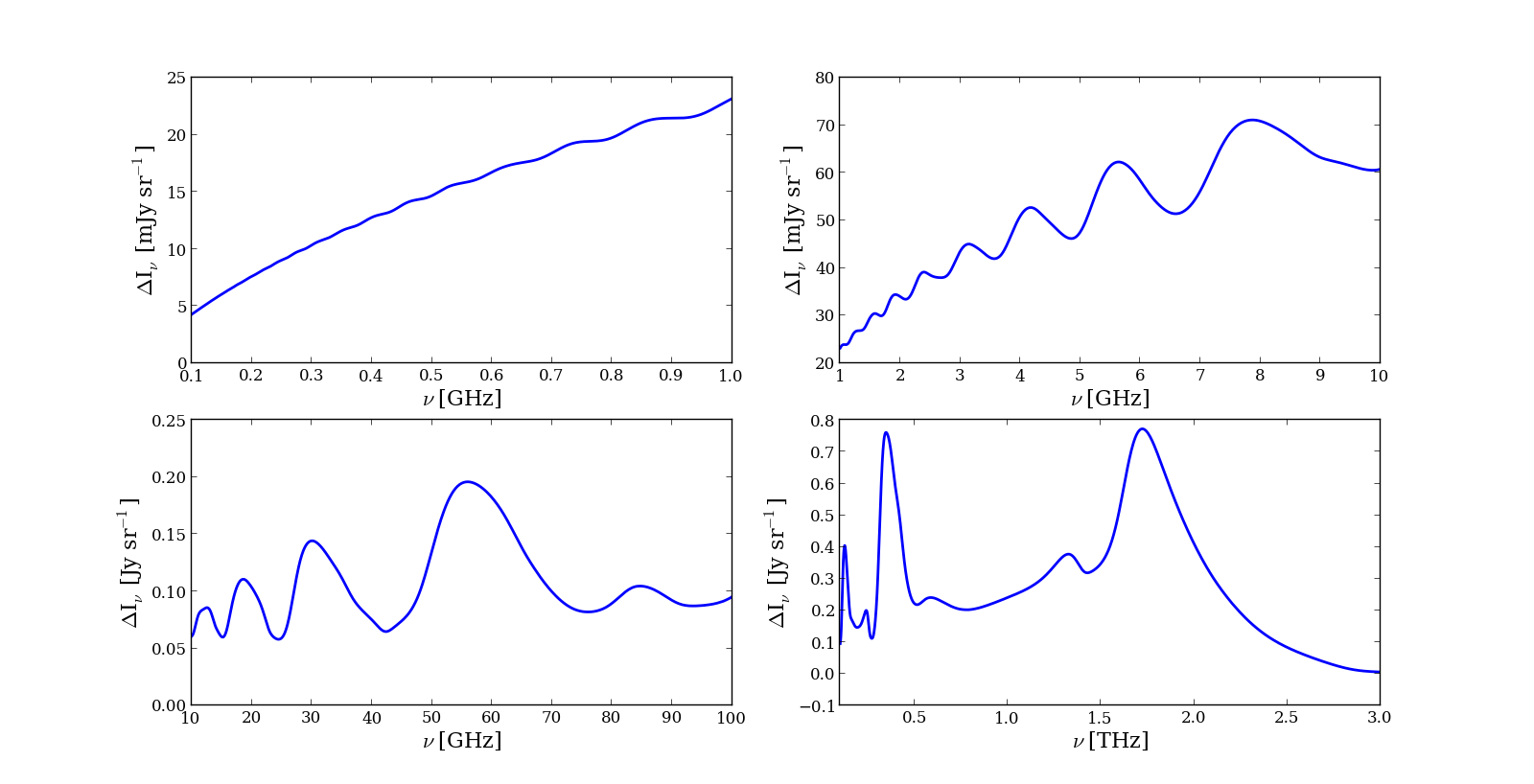} 
\caption{The additive spectral  structure expected in the intensity of the uniform extragalactic background light owing to cosmological recombination of Hydrogen and Helium \citep{Jose2006, Chluba2006b, Chluba2007, Jose2008}. The wide frequency range is successively covered in four panels. The additive spectrum is in units of [mJy\ sr$^{-1}$] in the panels in the top row and [Jy\ sr$^{-1}$] in the bottom row panels.}
\label{fig:template}
\end{figure}

In Fig.~\ref{fig:template}, we show the predictions from 0.1~GHz up to 3000~GHz. At high frequencies we can distinguish features caused by the Lyman-$\alpha$ line, Balmer-continuum and-$\alpha$ line, Paschen-$\alpha$ and Brackett-$\alpha$ lines \citep{Jose2006, Chluba2006, Chluba2007}. Helium contributes locally at a typical level\footnote{Enhancements of features caused by additional feedback effects \citep{Chluba2009c} were neglected for now.} of $\simeq 10\% - 30\%$ \citep{Jose2008, Sunyaev2009}. Owing to the substantial line broadening resulting from the extended period of recombination, at low frequencies the recombination radiation appears as (i) a smooth continuum to the radio, microwave and near IR backgrounds, plus (ii) more easily distinguishable spectral ripples that has a quasi-periodic frequency dependence.  It is the detection of these ripply features that we focus on in this work because of its distinctive signature that distinguishes it from any other known spectral features in the CMB. 

\subsection{The importance of an experimental detection of the recombination radiation}

Direct observations of the predicted ripples -- the unique {\it fingerprint} of the recombination era -- would be a confirmation of the recombination theory, validating our understanding of the associated atomic physics as well as the physical processes in these early times, which in itself is a major motivating factor for experimentally detecting them. A detailed discussion on the dependence of the spectral signatures from the epoch of recombination on various contributing factors that affect their strength, shape and position is presented in \citet{Chluba2008T0} and \cite{Sunyaev2009}. We list here some of the major motivations for an experimental detection.

\begin{enumerate}
\item {\it Determining key cosmological parameters.} Observing the spectral distortions from the epochs of Hydrogen and Helium recombination would, in principle, provide an additional way to determine some of the key parameters of the Universe. For instance, the dependence of the predicted recombination spectrum on $\Omega_{\rm b} h^2$ and $T_0$  is shown in Fig.~3 and 4 of \citet{Chluba2008T0}. It may be noted here that most of the recombination of Hydrogen and Helium and thus the spectral signatures that arise from these processes occur over redshifts {\it before} the formation of the CMB anisotropies that occurred near the peak of the Thomson visibility function: spectral lines from transitions associated with the Hydrogen atoms arise from a redshift range of 1300--1400. Hence detections of the recombination lines with increasing accuracy provides a way of constraining the thermal evolution of the universe beyond the last scattering surface (see below). This also provides a novel method to measure the {\it pre-stellar abundance} of Helium and to break parameter degeneracies by combining with CMB anisotropy measurements. An illustration of the dependence of the predicted final recombination spectrum on contribution from HeIII $\rightarrow$ HeII and HeII $\rightarrow$ HeI is given in Fig.~11 in \cite{Sunyaev2009}. 

\item {\it Probing energy release in the pre-recombination era.} Detection of cosmological recombination lines allows us to gain a better understanding of the thermal history of our Universe on the basis of the different redshifts of Hydrogen and Helium recombination. For instance, it is possible to compute the recombination spectrum assuming that the ambient radiation field is a distorted blackbody, where the distortion is of $y$-type \citep{Chluba2008c}. The $y$-distortion  \citep{Zeldovich1969} could arise from the blue-shifting of photons by Comptonization in the early universe, due to energy-release at $z\lesssim 50000$, when the redistribution of photons in energy by Compton scattering becomes inefficient \citep{Burigana1991, Hu1993, Chluba2011therm}. This distortion is described by the Compton-$y$ parameter, $y = \int \frac{k (T_{\rm e}-T_\gamma)}{m_{\rm e} c^2}\sigma N_{\rm e} c \,{\rm d}t$. COBE/FIRAS observations place an upper limit of $y< 1.5\times 10^{-5}$ \citep[95\% c.l.;][]{Fixsen1996, Fixsen2009}. The contributions from Hydrogen and Helium to the total recombination spectrum depends both on the value of $y$ parameter and when the distortion was created \citep{Chluba2008c}. Thus the cosmological recombination radiation may allow distinguishing between Compton $y$-distortions that were caused by energy release before or after the epoch of recombination. 

\item  {\it Testing recombination physics}. Todays most advanced recombination codes \citep{Chluba2010b, Yacine2010c} include many subtle atomic physics and radiative transfer effects \citep[e.g.,][]{Dubrovich2005, Chluba2006, Kholupenko2007, Switzer2007II, Chluba2008a, Hirata2008, Grin2009}. These calculations ensure that the science return from {\it Planck} and upcoming CMB experiments is not compromised by inaccuracies in the recombination model \citep{Jose2010, Shaw2011}. However, the detailed dynamics of the recombination process is also reflected in the shape and position of the recombination features. Any departures of the recombination spectrum from the theoretical predictions will indicate presence of some non-standard process \citep[e.g.,][]{Chluba2008c, Chluba2010a}. Thus, by observing the recombination radiation we can directly confront our understanding of recombination physics with experimental evidence. 

\end{enumerate}

\noindent Detecting nK amplitude fluctuations in the extragalactic background brightness, which are indeed a tiny perturbation to the orders of magnitude larger sky brightness temperature that arises from the CMB, extragalactic sources and Galactic emission, is a very challenging problem. However, we opine that receiver technology and the ability to produce detector arrays have progressed to the point where it is meaningful to look at the practical issues.  In this first paper, we study the feasibility of such a detection by modeling the sky spectrum as recorded by an ideal instrument and examine whether it is at all possible to recover the weak signal embedded in the substantially brighter Galactic and extragalactic foregrounds.  We discuss methods for fitting to data for the recovery or detection of the faint recombination line spectrum when observed embedded in the substantially brighter cosmic radio background. We also compute optimal observing frequencies for the detection and associated signal-to-noise ratio for detection with realistic receivers and a purpose-built array of spectral radiometers with due consideration to contribution from atmospheric emission further guided by radio frequency interference (RFI) over the band. Some of the challenges are very similar to those for the detection of the global 21cm signal \citep[see][]{Patra2013}; however, the recombination ripples benefit from their unique frequency dependence, which is hard to mimic by other sources and instrumental effects.

\section{The signal and noise level for a detection of the recombination radiation}\label{sec:snr}

\begin{figure}[ht]
\centering
\includegraphics[trim= 30mm 0mm 20mm 10mm, clip,scale = 0.4]{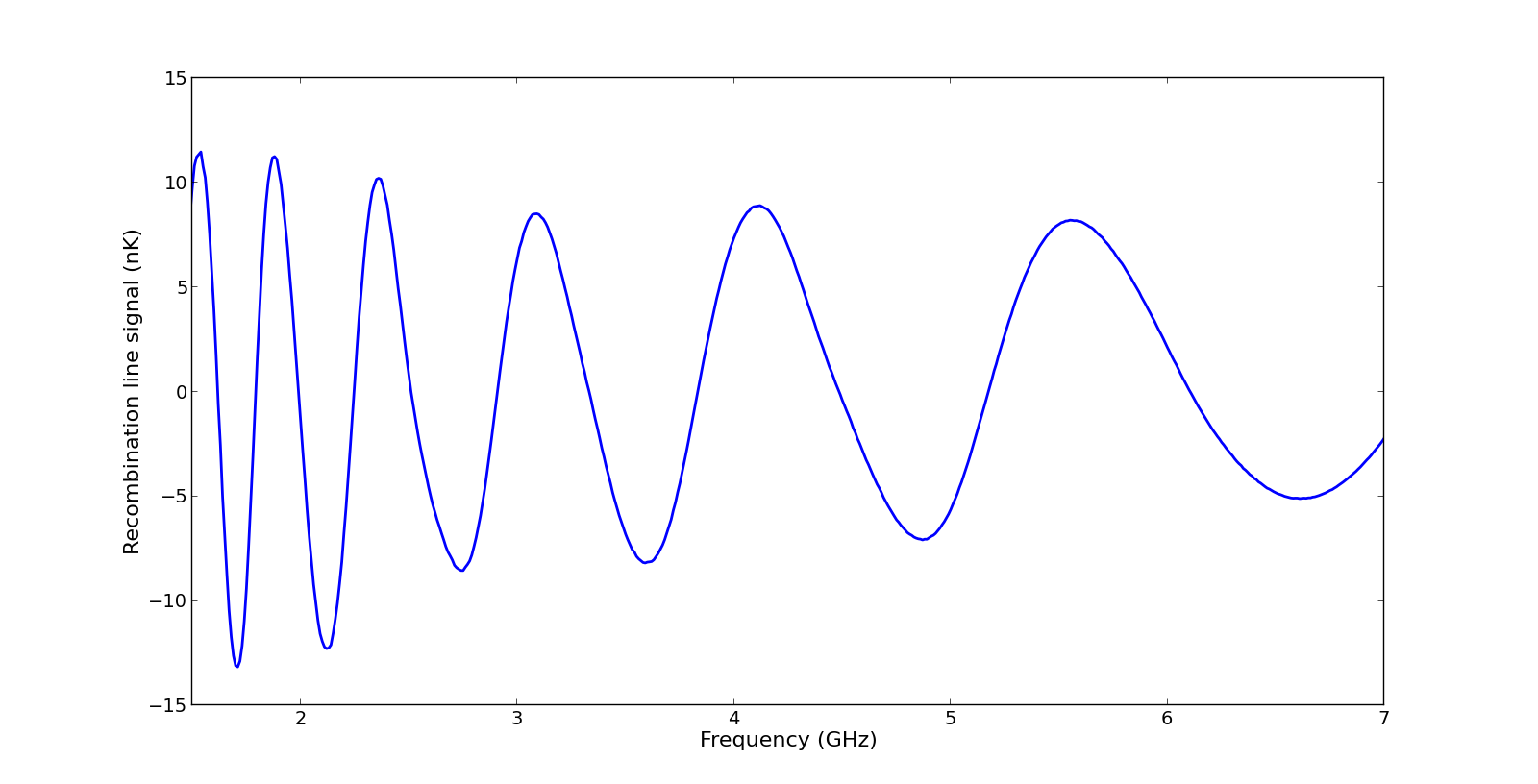}
\caption{ The ripply recombination line signal expected to be detected in the 1.5--7.0 GHz band.}
\label{fig:template_ideal}
\end{figure}

In this section, we discuss the frequency dependence of the spectral signal and that of the background brightness and additive instrument noise to arrive at estimates for signal-to-noise ratio versus frequency for detection of the line structure. The successive peaks of the recombination spectral features correspond to the spectral lines arising from bound-bound $n \rightarrow (n-1)$ electron transitions between adjacent principal quantum states of Hydrogen, visible at frequency $\nu \simeq 4.7\,\GHz\, [n/10]^{-3}$ (for emission redshift $z\simeq 1400$).  If we subtract a low order baseline component  from the recombination spectrum in the 1.5--7.0 GHz band, the resulting spectral structure is as shown in Fig.~\ref{fig:template_ideal}.  As a method of removing the foreground and receiver response in an observation with an ideal instrument,  if a smooth baseline were to be subtracted from a recorded sky spectrum,  Fig.~\ref{fig:template_ideal} is what would be expected as a residual. 

We now address the problem of identifying the most suitable frequency range for the experimental detection of recombination lines from the Epoch of Recombination by first considering the relative amplitude of the spectral ripples arising from recombination to the noise in the detection.  The total system noise is the additive sum of the spatially-varying sky background, consisting of contributions from discrete and diffuse Galactic and extragalactic sources in the sky, the cosmic microwave background and the cosmic infra-red background, as well as effects of atmospheric emission, plus receiver noise that is at lowest quantum noise associated with zero-point fluctuations. This sets a fundamental limit on the achievable detection sensitivity. Later in this section we also consider detection using receivers that have more realistic noise temperatures consistent with current technology.  We also discuss constraints arising from the requirement that within the observing band we need to cover at least a distinctive segment of the quasi-periodic cosmological signal, which might provide a unique template specific to cosmological recombination that would be difficult to mimic by other astrophysical sources, atmospheric emission, RFI or instrumental effects. 

\begin{figure}[Ht]
\centering
\includegraphics[ trim= 75mm 0mm 50mm 10mm, clip,scale = 0.5]{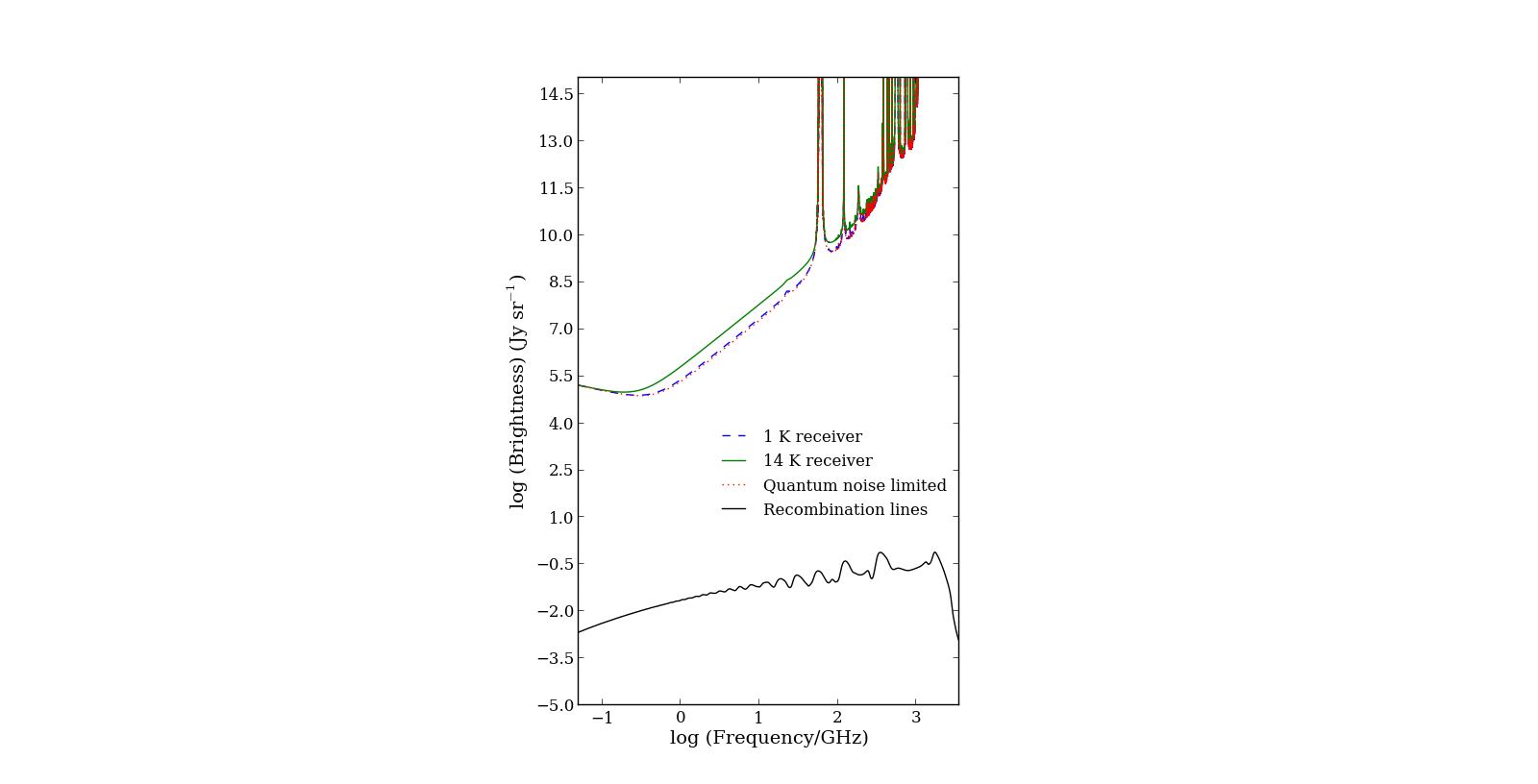} 
\caption{The expected intensity of the recombination signal is shown in black.  Our model for the intensity equivalent to the minimal total system noise is shown as a red dotted line where the receiver noise is assumed to have ideal quantum noise limited performance. Also shown as a blue dashed line is the intensity corresponding to system noise when observing with a state of the art cryogenically cooled receiver that has a noise temperature of 1 K and in green is the sky intensity when observing with an uncooled receiver assuming a noise temperature of 14 K. All intensities are in units of Jy~sr$^{-1}$ (surface brightness units) with the x-axis in $\log(\nu/\GHz)$.} 
\label{fig:lines_vs_sky_tcmb_trec}
\end{figure}

As a representation of the telescope system noise contribution from additive sky radiation, we construct a model spectrum of the background sky brightness all the way from 50~MHz up to 4~THz, extending to just beyond the redshifted Lyman-$\alpha$ line that is essentially the highest frequency at which we may expect spectral features arising from cosmological recombination.  We adopt the model presented in \cite{Subrahmanyan2013} for the Galactic and extragalactic contributions to sky brightness and derive sky temperatures towards the Galactic pole at 150, 408 and 1420~MHz.  A fit of a power-law form to these brightness estimates yields a temperature spectral index of $-2.5$ and a normalization corresponding to sky brightness of 438~K at 100~MHz; this model spectrum towards the Galactic pole is adopted to represent the contribution to system noise from Galactic emission plus extragalactic discrete sources.  We also include a component that is an estimate of the far and mid-infrared background; this model is derived from data in Table~47 of \citet{Leinert1998}.   For the receiver noise, we compute the system temperature for two cases: (i) assuming a state of the art cryogenically cooled receiver with noise temperature of 1~K above quantum noise \citep{Schleeh2012} and (ii) that for an uncooled receiver with a more realistic noise temperature of 14~K plus quantum noise \citep{Belostotski2007,Witvers2010}.  We use the {\it am}  atmospheric model \citep{Paine2004} with typical conditions at the Chajnantor\footnote{https://www.cfa.harvard.edu/\~spaine/am/cookbook/unix/other\_examples/Chajnantor.amc} plateau site to derive the atmospheric opacity versus frequency.  The system temperature is finally translated to above atmosphere using this opacity so that it may be combined with the signal intensity derived above for calculating the signal-to-noise ratio. Our adopted model spectrum for the above-atmosphere system temperature  is shown in Fig.~\ref{fig:lines_vs_sky_tcmb_trec} along with the spectrum of the expected recombination signal. 

\begin{figure}[h]
\centering
\includegraphics[trim=10mm 0mm 10mm 0mm,clip,scale = 0.5]{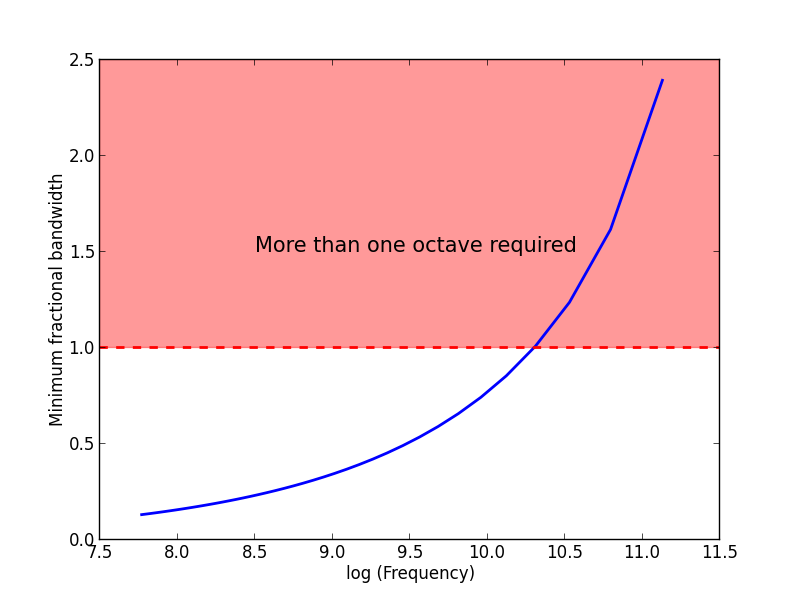} 
\caption{Ratio of spacing between peaks of $\alpha$ transitions from $n$ and $n+2$ quantum shells of Hydrogen to the nominal frequency versus nominal observing frequency.} 
\label{fig:bw_vs_nu}
\end{figure}

As a criterion for the minimum bandwidth required at any observing frequency we choose a spectral window that includes at least two adjacent broad recombination spectral lines that arise from $n \rightarrow (n-1)$ ($\alpha$) transitions.  This minimum bandwidth would correspond to the spacing between peaks of $\alpha$ transitions from $n$ and $n+2$ shells.  Such a spectral window would be expected to include a sufficiently high order of the variation in the signal so as to give it a distinctive signature.  Fig.~\ref{fig:bw_vs_nu} shows the ratio of this minimum bandwidth to the nominal center frequency versus the center frequency. The detection of Recombination Epoch spectral lines clearly requires a larger fractional bandwidth at higher frequencies.  A ratio less than unity implies that an octave bandwidth at the observing frequency would contain more than two adjacent recombination lines, satisfying our criterion.  The ratio falls below unity for observing frequencies below about 18~GHz, which implies that detecting the cosmological line spectrum would require bandwidths exceeding an octave if the center frequency were to exceed 18~GHz.   Receivers with bandwidths wider than an octave are susceptible to self-generated radio frequency interference from harmonics of system clocks and local oscillator frequencies; therefore, although sensitivity for a detection would undoubtedly improve with wider observing bands, for the ultra-sensitive detection experiment being considered herein it would be unwise to attempt detection at frequencies above this value. 

\begin{figure}[ht]
\centering
\includegraphics[trim=30mm 0mm 30mm 0mm,clip,scale = 0.40]{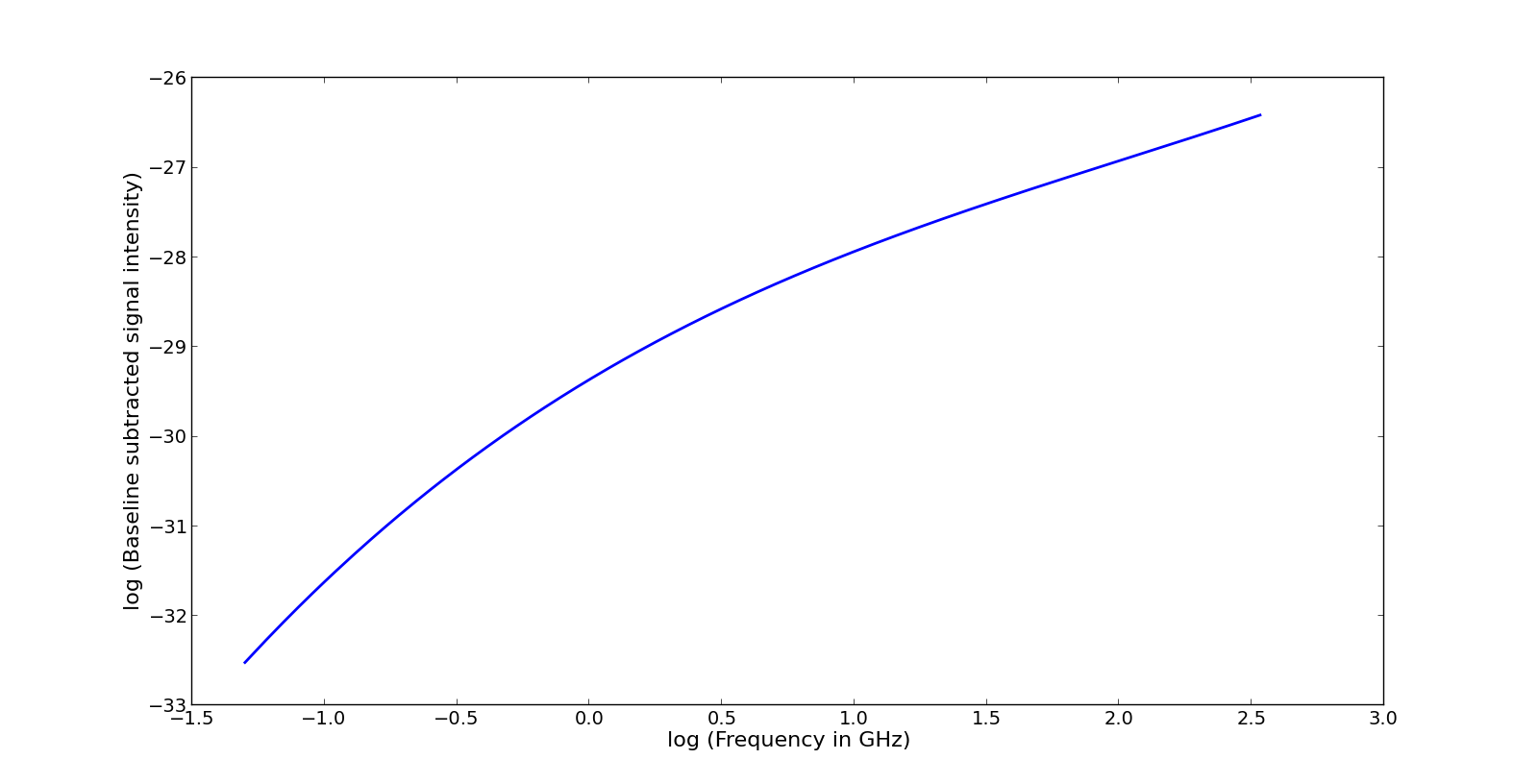}
\caption{Variation in the logarithm of amplitude of the ripples representing the signal from recombination versus frequency in log(GHz) units.}
\label{fig:sig_cont}
\end{figure}

As an estimate of the detectable signal from Hydrogen recombination, we require a measure of the peak-to-peak spectral variations versus observing frequency.  We first fit a spline through the points in the recombination line intensity template that are at frequencies corresponding to the peaks of the $n \rightarrow (n-1)$ Hydrogen transitions.  We use a redshift that is a mean value, which we determine by comparing the locations of the peaks in the predicted recombination line spectrum with the corresponding locations given by the Rydberg formula.  Subtracting the spline fit gives a residual recombination line spectrum that represents the detectable signal in an observation from which a mean spectral baseline has been subtracted.  In Fig.~\ref{fig:sig_cont} we plot the strength of the expected recombination line signal defined as half the peak-to-peak amplitude of the baseline subtracted ripples expected in the observed band, versus observing frequency.   With frequency, the expected ripple amplitude increases by orders of magnitude across the frequency range we consider here.  It may be noted that although in terms of spectral intensity this ripple signal amplitude increases with frequency, as shown above the number of spectral ripples within any octave bandwidth decreases with increasing frequency. Thus, a frequency regime that maximizes both aspects is sought. (See Sect.~\ref{sec:freq_choice})

\subsection{Signal-to-noise ratio: a matched filter approach to signal detection}

The detection of spectral lines from cosmological recombination involves first subtracting a baseline from the observed spectrum to remove the relatively smoother foregrounds and CMB, after which the residual spectral segment may be examined for the expected spectral ripple using a matched filter.  We evaluate here the signal-to-noise ratio for a detection method where an estimate of the amplitude of any spectral ripple that matches a theoretical template is derived from the measurement.   The expected signal, following baseline subtraction, is similar to a sinusoidal ripple that has a period that varies systematically across the spectral segment.  The noise or uncertainty in the estimate of the amplitude of the ripple is derived from the measurement errors in the measured intensities in the spectral channels, and from the propagation of errors from these measurements to the derived estimate of the amplitude of the signal present in the data.  The measurement errors in channel data depend on the system noise, which originates from the sky brightness and receiver noise, and on the channel bandwidth and integration time.

We consider here the ideal scenario where the measurement data---the bandpass calibrated spectrum of the cosmic radio background from which a baseline has been subtracted---contains a spectral ripple that is the same as the expected theoretical template without any other residual contamination.  In this case where the signal amplitude is estimated using a matched filter derived from the template, we obtain a signal-to-noise ratio that is limited purely by the noise present in the spectral channels. This case study admittedly represents an optimistic estimate of the signal-to-noise ratio attainable. 

The residual spectrum after baseline subtraction is assumed to have $N$ frequency channels, each of bandwidth $\delta b$. The sampled residual sinusoidal ripple, which is also the template of the expected ripply recombination lines following baseline subtraction, is given by $a_k$ and has an amplitude of $A_0$. From this template of the expected signal we define a weighting function $w_k$, which is defined such that the weighted sum of $a_k$ is $A_0$. The matched filter is thus the weighted summation of the observed spectrum and the matched filtering yields an estimate of the amplitude $S$ of the ripple from recombination:
\begin{equation}\label{eq:sum_aw}
S=\sum_{k=0}^{N}a_kw_k
\end{equation}
The template representing the expectation and the corresponding weighting function may be scaled appropriately to search the data for signatures of the ripple from recombination assuming a range of effective redshifts for the recombination lines.  

The weighting function is hence similar to the template in that it has the same ripple form but with an amplitude $w_0$.  If the signal matches the template, the requirement that $S=\sum_{k=0}^{N}a_kw_k$ ought to equal $A_0$ leads to the condition that $w_0 = 2/N$.

The uncertainty in the channel data in the residual spectrum $a_k$ is 
\begin{equation}
\delta a_k = \frac{T}{\sqrt{\delta b \cdot t}},
\end{equation} 
where $T$ is the system temperature and $t$ is the total integration time.  Propagating this error in the channel data to the error $\delta S$ in the estimate of the amplitude of the ripple from recombination: 
\begin{eqnarray}
\delta S & = & \sqrt{\mathlarger{\sum}_{k=0}^{N-1} (\delta a_k)^2 \abs{w_k}^2} \nonumber \\
 & = & \frac{T}{\sqrt{\delta b \cdot t}} \cdot \sqrt{\mathlarger{\sum}_{k=0}^{N-1} \abs{w_k}^2} \nonumber \\
 & = & T w_0 \sqrt{\frac{N}{2 \delta b \cdot t}  }. \nonumber
\end{eqnarray}
The last step above follows from the result that the average value of the square of a sinusoidal wave is half the amplitude.  Substituting the earlier result that $w_0 = 2/N$ yields:
\begin{equation}
\delta S = T \sqrt{\frac{2}{N \delta b \cdot t}}.
\end{equation}

Thus the signal-to-noise ratio in the measurement is given by:
\begin{equation}\label{eq:snr}
SNR = \frac{S}{\delta S}= \frac{A_0}{T}\sqrt\frac{N \delta b \cdot t}{2} = \frac{A_0}{T}\sqrt\frac{B \cdot t}{2},
\end{equation}
where $B = N \delta b$ is the total bandwidth of the observed spectrum. For this estimate it is assumed that noise in the different channels is uncorrelated.


\subsection{Choice of Frequency for detecting ripples from cosmological recombination}\label{sec:freq_choice}

\begin{figure}[ht]
\centering
\includegraphics[trim = 30mm 0mm 0mm 0mm, clip,scale = 0.45]{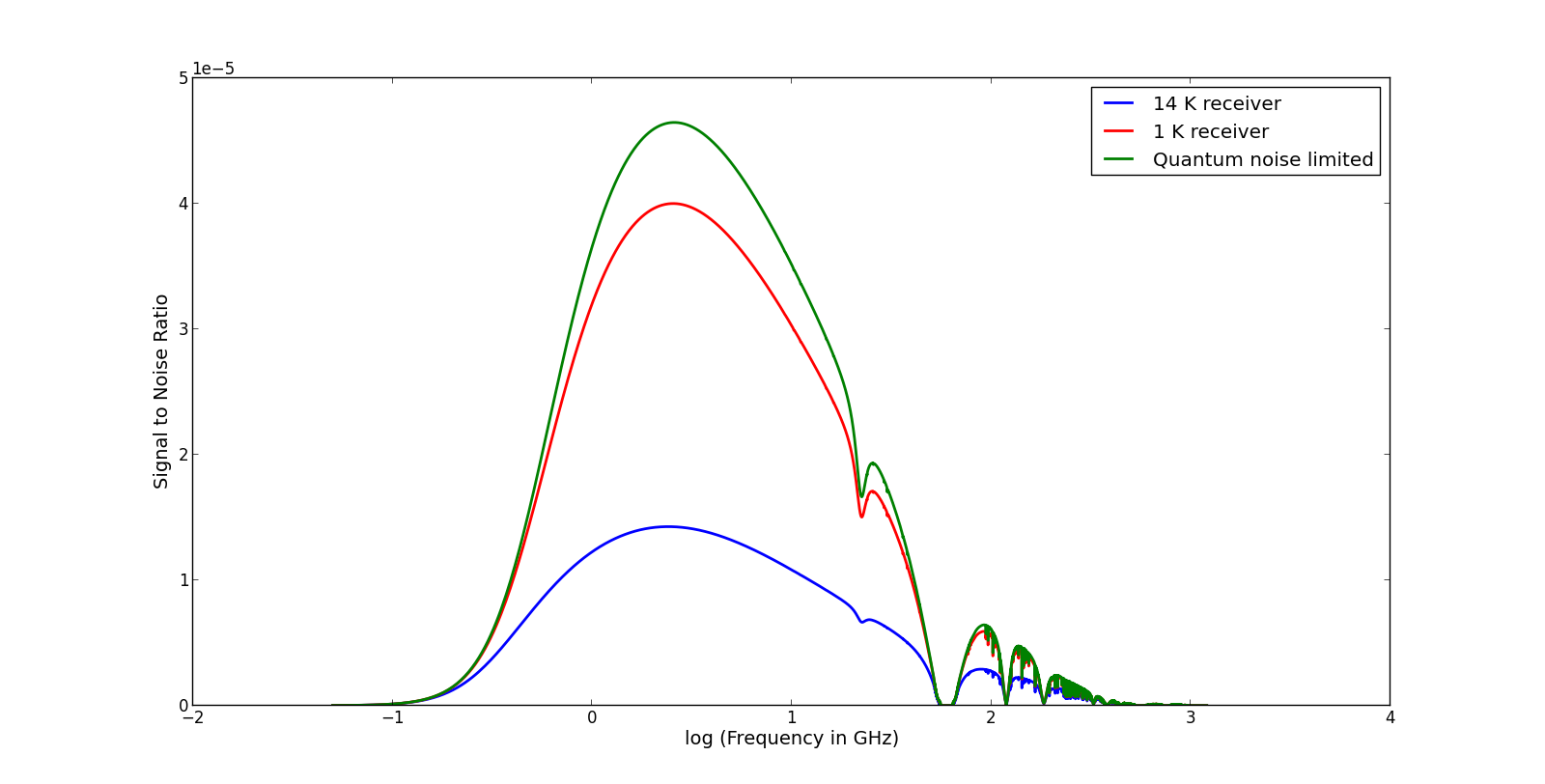} 
\caption{Signal-to-noise ratio versus nominal observing frequency (in log(GHz) scale) for the detection of spectral-line signatures from recombination epoch with an octave bandwidth. The green line shows the ratio assuming ideal quantum noise limited receivers; the blue line shows the ratio for the more realistic case of receivers with 14~K receiver noise. The red line shows this ratio for a receiver which is cryogenically cooled to 1~K. The observing time is assumed to be 1~s.} 
\label{fig:snr_cmb_rec}
\end{figure}

In Fig.~\ref{fig:snr_cmb_rec}  we show the estimated signal-to-noise ratio versus observing frequency to guide the choice of observing frequency. In creating this plot the signal is assumed to be half the peak-to-peak magnitude of the ripple, which is the amplitude shown in Fig.~\ref{fig:sig_cont}. The noise is assumed to be the system temperature (from Fig.~\ref{fig:lines_vs_sky_tcmb_trec}) scaled by a factor $\sqrt{2/B}$, where $B$ is the octave bandwidth about a nominal central frequency.  The integration time is assumed to be 1~s. The figure shows the signal-to-noise ratio for the case where the system noise corresponds to quantum-noise limited receivers (in green); we also show in red the signal-to-noise ratio for a  cryogenically cooled receiver with 1~K noise temperature and in blue this ratio for the case of uncooled 14~K receivers. All three traces are assuming an ideal case where the spectral ripple in the measurement data matches the predicted template exactly. 

Detection with maximum signal-to-noise ratio suggests observing in the 1--6~GHz band. If we avoid the lower end of this band where Galactic HI and terrestrial and satellite down-link related radio frequency interference is substantial, the 2--6~GHz band is suggested. Observing in octave bandwidths within this range also satisfies the criterion that at least two cycles of ripples of the recombination line spectrum ought to be contained in the observed spectral segment (see Fig.~\ref{fig:template_ideal}). 



\section{A modeling of observations of the cosmological recombination spectrum}

The recombination lines from cosmological recombination are detected by a receiver along with foregrounds, which are averaged by the telescope beam over its response pattern on the sky.  The averaging over a multitude of sources with different emission spectra, over the sky and along line of sight, results in a detected spectrum that would have an unknown form: even if the emissivity of the gas is a power-law form at every location, the spectral index does vary across the sky and along line of sight and, therefore, the averaging by the telescope beam would result in an observed spectrum that deviates from a single power law.  A similar effect, related to the superposition of blackbodies inside the beam of an experiment, causes an inevitable $y$-type distortion of the CMB spectrum \citep{Chluba2004}; the superposition of blackbodies simply is not a blackbody anymore \citep{Zeldovich1972, Chluba2004, Stebbins2007}.
It is necessary to detect the presence of the recombination line spectrum in the observed spectrum although the line spectrum has amplitude that is {\it almost nine orders of magnitude smaller} and the additive foreground is a complex spectrum of unknown form! By simulating the sky spectrum as would be observed by an ideal system, this section addresses the question of whether such a detection is at all possible.

We choose an octave band from 3 to 6~GHz as the observing band for the simulations presented here.  This band also meets our criterion of having at least two recombination spectral lines within the band.  In this section we describe a code we have developed, which simulates an ideal receiver system observing the sky over the identified frequency range, and generates the temperature spectrum of the sky as would be produced by such an instrument.   A calibration method is included in the pipeline.

We use all-sky maps at 408~MHz \citep{Haslam1982}, 1420~MHz \citep{Reich1982, Reich1986, Gorski2005} and at 23~GHz (WMAP science data product\footnote{WMAP Science Team}) as input maps to estimate the combined galactic and extragalactic brightness contribution to the sky spectrum between 3 and 6~GHz. The sky brightness in the Rayleigh-Jeans limit, at every sky pixel and over frequency channels spaced 10 MHz apart is derived from these three input temperature maps using a linear interpolation in log-log space. 

We derive a 408-MHz all sky map from the corresponding raw data product available in the LAMBDA\footnote{http://lambda.gsfc.nasa.gov/} website. The non-destriped 408-MHz map made with a $0\fdg85$ beam that is available in the HEALPix \citep{Gorski2005} R9 nested ordering scheme is first smoothed with a Gaussian beam of FWHM $31\farcm60$ to obtain a map with $1^{\circ}$ resolution. Following this we degrade the map to an R8 nested ordering representation. This final 408-MHz map used in the simulation is a nested R8 HEALPix map in Galactic coordinates with a resolution corresponding to beam FWHM $1^{\circ}$ and with temperature in mK brightness units. 

The 1420-MHz all sky map was derived from the data available in the OCEANCOLOR\footnote{http://oceancolor.gsfc.nasa.gov/AQUARIUS/DinnatEtAl2010/} website. This data is in Kelvin units with a beam size of $0\fdg6$ and a pixel size $0\fdg25$ and is in J2000 celestial coordinates. We perform gridding convolution of the data with a Gaussian of appropriate FWHM to yield an image with final resolution of $1^{\circ}$. We then transform the map to Galactic coordinates using the celestial HEALPix coordinate pixel listing for the R8 nested ordering scheme. Thus the final all-sky map we use at 1420 MHz has a beam FWHM $1^{\circ}$ and is in Galactic coordinates ordered in the nested R8 HEALPix scheme.

The deconvolved form of the WMAP 23~GHz all sky map from the LAMBDA website was smoothed to $1^{\circ}$ resolution and formatted in Galactic coordinates in the HEALPix R8 nested ordering scheme. The image units were transformed from thermodynamic to brightness temperature.  The image intensities correspond to a differential measurement; therefore, the map is uncertain in its zero point, which was arbitrarily set.  We assume a plane-parallel slab model for the galaxy and use the method adopted in \citet{Kogut2011} to estimate this uniform component at 23~GHz using all sky images of the absolute brightness of the sky at 150, 408 and 1420~MHz.  The CMB monopole, dipole and WMAP ILC images were subtracted from all these images and the pixels with substantial contamination from discrete sources were blanked using the same blanking image that was used in the WMAP analysis. Pixels in each of the images were binned in cosecant Galactic latitude (cosecant($b$)) over the range 1.0--4.0.  The run of median pixel intensity versus mean cosecant($b$) were examined in log-log coordinates. A straight line fit to this plot yields an estimate of the mean extragalactic background at each frequency. It may be noted here that we have restricted this analysis to the southern Galactic hemisphere for consistency with the WMAP analysis. For the case of the 23~GHz image, the fit yields an intercept of $0$K with an accuracy of 1 part in $10^9$ as expected for an image has been constructed to have its intercept arbitrarily set to zero.  The intercepts estimated for the lower frequency images were extrapolated using a polynomial fit to get the `missing' uniform background in the 23~GHz image. The resulting value of $493~\mu$K was added as a constant to the 23~GHz image.

The three maps representing the Galactic and extragalactic radio emission (minus CMB) at 408~MHz, 1420~MHz and 23~GHz  were interpolated at every image pixel separately to estimate sky temperatures at frequencies between 3 and 6~GHz.  We interpolate using a first order polynomial in log-brightness-temperature versus log-frequency space, considering the spectrum at every pixel to be a smooth power-law in linear space. The total sky brightness toward any sky pixel and frequency is estimated as the sum of the galactic and extragalactic sky brightness derived from this polynomial interpolation, plus a uniform cosmic microwave background brightness computed using the Planck formula, plus a uniform component corresponding to the weak cosmological recombination line spectrum, as shown in Fig.~\ref{fig:template}.

\begin{figure}[Ht]
\centering
\includegraphics[scale = 0.55]{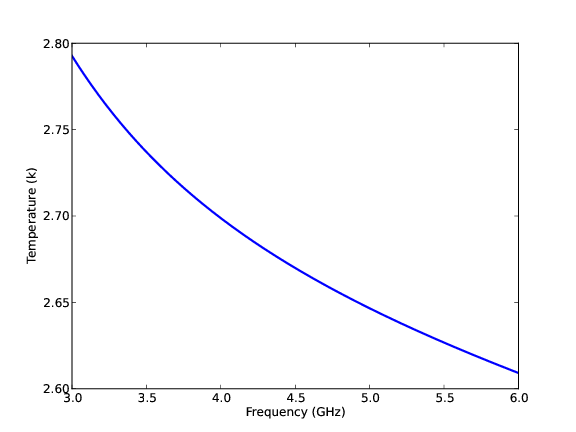} 
\caption{ Sample synthetic spectrum recorded by an ideal receiver system discussed in the text.  The receiver is assumed to be observing with an antenna with a cos$^2(ZA)$ beam pattern pointed at the zenith.
}
\label{fig:sample_spectrum}
\end{figure}

\begin{figure}[Ht]
\centering
\includegraphics[trim=10mm 0mm 0mm 0mm,clip,scale = 0.5]{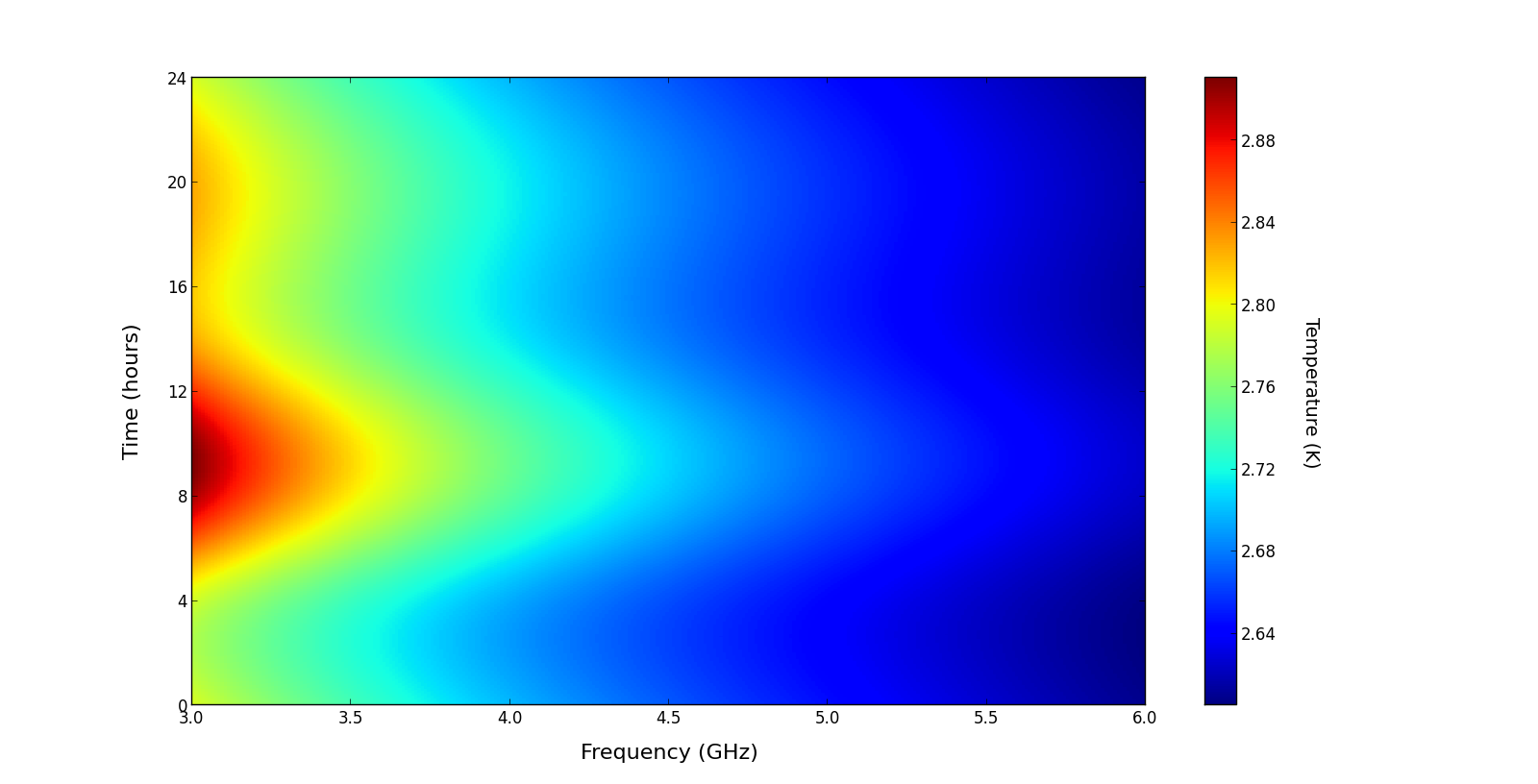} 
\caption{ A time-frequency plot showing the synthetic sky spectra as observed over 24~hours. One can see the galaxy rise and set as the sky drifts over the instrument with each spectrum recorded one minute apart.}
\label{fig:spectrum_24hr}
\end{figure}

The observing system is assumed to have a single antenna element with a frequency independent radiation pattern that has a cos$^2(ZA)$ form, where $ZA$ denotes zenith angle.   The antenna bore sight (the optical axis the antenna) is assumed to be static on the ground at the observing site and directed towards zenith at all times. We neglect polarization in the sky intensity as well as the telescope response. Separately, the spectral radiometer is assumed to be a correlation spectrometer and hence does not respond to receiver noise \citep[see][]{Patra2013}.  The spectral response of the observing system at any instant is then simply the weighted sum of spectra towards every pixel in the sky where the weighting is by the beam power pattern towards that sky direction.  As stated above, the spectrum towards any sky pixel is not expected to be a simple power law because it is an average along the line of sight of regions that might have different spectral indices. The weighted average spectrum is an average over the sky and along the line of sight and hence even if each emitting region is of power law form the average observed spectrum may have a complex form that is unknown. 

Our simulation code has the geographic location of the observing site and LST as free parameters. Our code can be easily modified to change the antenna beam pattern as well.
The generation of spectra takes into account effects of atmospheric refraction and precession;  astronomical aberration is negligible even in its severest form for the problem considered here  and is dropped from our calculations.   Calibration is assumed to be done by recording spectra with a hot (373.0~K) and separately a cold (273.0~K) load on the antenna and dividing spectra recorded on the sky by the difference between spectra recorded with the hot and cold loads. Bandpass calibration of sufficiently high precision is assumed. Our code generates mock calibrated spectra over time whose mean temperature varies as the sky and Galactic Plane drift across the telescope beam. A sample spectrum of the sky is shown in Fig.~\ref{fig:sample_spectrum}.
Figure~\ref{fig:spectrum_24hr} shows a waterfall plot to illustrate the variation over 24~hours in the recorded spectra observed by this ideal instrument as the sky drifts across the antenna.   

The synthetic sky spectra, whose amplitude is of the order of a few K, contains the recombination line spectrum as a small additive component and is representative of mock observations made with a correlation spectral radiometer  that does not respond to receiver noise.  The challenge is to distinguish the weak signal corresponding to the epoch of cosmological recombination, which is buried in the observed spectrum as a broadband quasi-periodic sinusoid with peak-to-peak amplitude of order $\sim10$~nK. {\it Not only is the signal a tiny fraction of the total sky spectrum, the recombination line spectrum is an additive component of a foreground spectrum whose functional form is complex and unknown.}  In the next section we discuss methods to detect the cosmological recombination spectrum and challenges therein.

\section{The detection of signatures of cosmological recombination}
\label{sec:fitting}
In the previous section we have described the generation of a synthetic spectrum of the radio sky between 3 and 6~GHz, as would be observed by an ideal instrument. The signal from cosmological recombination is expected to appear as quasi-periodic ripples, some $9$ orders of magnitude smaller than the galactic and extragalactic foreground spectrum in which it is additively concealed. The discernment of the recombination signal in such a total spectrum, even under ideal conditions, is pivotal in answering the question of whether an experimental detection is indeed possible. Below we discuss the challenges involved in and describe a possible method for the detection of the cosmological recombination signal in synthetic sky spectra.  

The brightness of the sky as measured by a total power radiometer is the sum of the brightness contributions from all the discrete and diffuse Galactic and extragalactic sources that lie within its beam, along the line of sight and across the sky, including the cosmic microwave background and other cosmological emissions. The precision with which the foreground needs to be modeled so that a subtraction of the model might reveal the cosmological recombination spectrum is extreme.  Although the discrete sources lying in the beam might have well-measured spectra and the diffuse sky radiation might have all-sky maps at multiple frequencies, the functional form of the final cumulative spectrum of the foreground is not known {\it a priori} to the required accuracy. The foreground has necessarily to be fit to the measurement, which implies that the functional form used for the fit to the foreground needs to be of a form that accurately fits to the foreground without also fitting to the recombination spectral features.  Only then may a fitting of a foreground model to the measurement set and its subsequent subtraction reveal the recombination line features in the residual.

The thermal and non-thermal processes (e.g., synchrotron and free-free emission) that contribute to the foregrounds have spectra that are smooth because they arise from impulsive emissions in time domain.  The electron energy spectra in astrophysical objects and diffuse matter are also believed to be smooth over decades in energy space.  Hence the cumulative spectra in thermal and non-thermal emissions are expected to be featureless over octave bandwidths and, therefore, ought to be distinguishable in principle from the cosmological recombination radiation.  

The problem of recovering broad and weak spectral deviations that are buried in a total sky spectrum that is several order of magnitudes brighter is not unique to cosmological hydrogen recombination lines. All-sky (also referred to as global) spectral deviations predicted to arise from the Epoch of Reionization (EoR) also face a similar challenge in that broad spectral features of about 10--100~mK brightness are buried in foregrounds of several hundred K, the foreground being $\sim5$ orders of magnitude brighter.Thus the importance of a method to effectively fit a smooth functional form to the combined Galactic and extragalactic foreground over 1-2 octaves of bandwidth, in which the signal of interest is itself not lost in the process, cannot be understated.  

\begin{figure}[Ht]
\centering
\includegraphics[scale = 0.5]{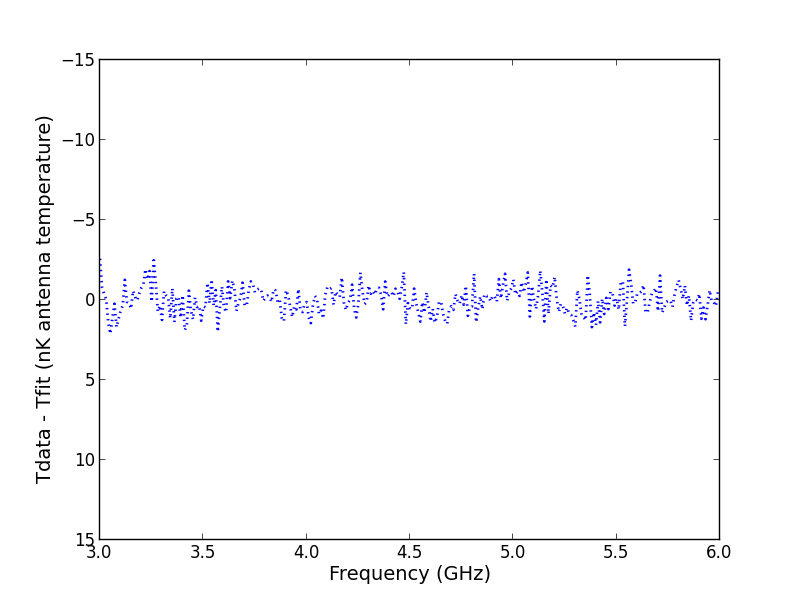} 
\caption{ The residual obtained on subtracting a sample sky spectrum by an eight order polynomial fitting function in log-frequency and log-temperature space. While one would expect to see signatures of the spectral lines arising from cosmological recombination in the residual at the nK level, the residual is dominated only by thermal noise with the lines themselves having been absorbed by the fitting function.}
\label{fig:unconstrained_residual}
\end{figure}

In the context of detecting global EoR signal in the all-sky radio background spectrum, a polynomial functional form of high order has been adopted in the literature as the analytic function to fit to the foreground \citep{Bowman2010}.  However, broad spectral lines of interest that are present in observed spectra as additive components may also be absorbed in the polynomial fit, particularly if the polynomial is of high order, such that the residual would have little trace of the cosmological signal.   As an illustration, on fitting a sample spectrum generated by our simulation with an eighth-order polynomial in log-frequency versus log-temperature space without any constraints on the nature of the polynomial, and subtracting the fit spectrum from the original spectrum, we obtain a residual as shown in Fig.~\ref{fig:unconstrained_residual}.  While one might hope that the polynomial fit to the foreground would leave the recombination line signal untouched, comparison with Fig.~\ref{fig:template_ideal} shows clearly that this is not the case.   The ripples from cosmological recombination, the very signal that we are looking to detect, has been eliminated by the fitting process. 

What is needed is a careful choice of the functional form that would leave the cosmological signals of interest almost wholly in the residue.  Also needed is a method of successive approximation that would model the foreground to the required accuracy so that the faint cosmological signal might dominate the residue.

\subsection{A complete monotone approach to foreground modeling}

The CMB component has a well defined functional form.  The fitting function $f_{\rm fg}(\nu)$ for the foreground may be defined to be an analytic function over the frequency range of interest, so that the model does not admit discontinuities and its derivatives are defined.  The functional form describing the foreground model is expected to be `smooth' in the sense that it {\it must not be able to} fit to additive signals that have the sinusoidal structure expected of the recombination spectrum.  

 A potentially useful functional form is that of a {\it completely monotonic} function.  Mathematically, a function $f(x)$ is said to be a complete monotone if for all values of $x$ in the interval $0 \le x < \infty$,  $(-1)^n\times\frac{\textrm{d}^nf(x)}{\textrm{d}x^n} \ge 0$ for every integer $n \ge 0$.   An example of a complete monotone function is $f(x)=1/(a+b x)^c$, where $a \ge 0$, $b \ge 0$ and $c \ge 0$.  This implies that power-law form spectra with negative spectral indices are completely monotonic functions.  It is also known that if $f(x)$ and $g(x)$ are completely monotonic, then $a f(x) + b g(x)$ where $a$ and $b$ are non-negative  constants is also completely monotonic,  which implies that the sum of power law spectra with negative indices would also be completely monotonic.  A function $f(x)$  may also be completely monotonic over finite range $a < x < b$ where $a \ge 0$ and $b \ge 0$: this would be the case if the condition $(-1)^n\times\frac{\textrm{d}^nf(x)}{\textrm{d}x^n} \ge 0$ for every integer $n \ge 0$ holds over the range $a < x < b$.  The mathematical definition suggests that if $f_{\rm fg}(\nu)$, where $\nu$ is frequency,  is the foreground model over a certain bandwidth, then we may adopt a functional form for $f_{\rm fg}(\nu)$ that is a complete monotone within the bandwidth of interest.

Successive approximation of an analytic functional form to data may be done using a Taylor approximation. The Taylor series of an analytic function always converges about every point in its domain.   We may represent $f_{\rm fg}(\nu)$ as a polynomial whose coefficients are constrained so that $f_{\rm fg}(\nu)$ is completely monotonic and expand the function as a Taylor polynomial by successively estimating the coefficients of the polynomial, stopping when the degree of the polynomial is such that it is a sufficiently good fit to the data ({\it i.e.} the observed sky spectrum) and the residual is dominated by the embedded cosmological recombination lines. It may be noted here that increasing further the order of the polynomial, whose coefficients are constrained such that the polynomial is a complete monotone, will only yield vanishingly smaller coefficients for the higher order terms. Since the constraint on the polynomial forces it to remain smooth, once the recombination lines dominate the residual the residual is not minimized further by the introduction of higher order terms. Thus we converge to a smooth completely monotonic approximation to the foreground that leaves the recombination line spectral structure as a residual to the fit, which no longer changes significantly with increasing the order of the completely monotonic polynomial.

$f_{\rm fg}(\nu)$ may be modeled as a completely monotonic (CM) polynomial in brightness temperature versus frequency space.  More useful is a modeling in temperature versus log-frequency space, where the number of terms in the Taylor approximation would be reduced.  The Taylor expansion may be conveniently computed using the lowest frequency in the range of interest as the reference value.

A CM polynomial of arbitrary order $n$, whose constant term $a_0$ is left unconstrained so as to allow arbitrary vertical translations, may be written in the form
\begin{equation}\label{eq:CM_function}
f(x) = a_0 + \sum_{i=1}^{n} (-1)^i (x - x_0)^i \{ \sum_{j=0}^{n-i} a_{i+j} C^{i+j}_{j} (x_m - x_0)^{j} \},
\end{equation}
where $C^n_k$ denotes the binomial coefficient $n!/\{k! (n-k)!\}$. An example illustrating the algorithm to construct such a CM function and thereby arriving at the functional form presented in equation \ref{eq:CM_function} is described in Appendix \ref{sec:appendix_a}.

A disadvantage of adopting a CM functional form is that the data is fitted in brightness temperature versus log-frequency space, where the foreground is expected to be CM.  A functional form that is smooth in log-temperature versus log-frequency space is preferred, since a lower order polynomial would be sufficient to describe the foreground.  Lower numbers of parameters makes for more robust fitting with less likelihood that optimization algorithms get trapped in local minima.  


\subsection{Modeling the foreground as a {\it Maximally Smooth} function:  a smooth polynomial that has no zero-crossings in derivatives}

We now consider polynomial functional forms to describe the smooth foreground in log-temperature versus log-frequency space. The first approximation in this parametrization is a straight line representing the mean spectral index of the sky region.  Curvature in the mean spectrum may be represented, to lowest order, by adopting a parabolic form for the spectrum in this log space.  This form has a constant second order derivative.  

If we improve upon the modeling of the spectral curvature by adding a cubic term to the polynomial, we may constrain the model to be smooth and without inflections by requiring that the second derivative has no zero crossings in the domain.  If we considering a polynomial model of the form
\begin{equation}
f(x) = p_0 + p_1 (x -x_0) + p_2 (x - x_0)^2 + p_3 (x - x_0)^3,
\end{equation}
we would require that second derivative
\begin{equation}
\frac{\text{d}^2f(x)}{\text{d}x^2} = (2!/0!) p_2 + (3!/1!)p_3 (x - x_0)
\end{equation}
has no zero crossings in the domain.

If we wish to improve, further, the modeling of the spectral curvature we may represent the foreground spectrum by a fourth order polynomial of the form:
\begin{equation}
f(x) = p_0 + p_1 (x -x_0) + p_2 (x - x_0)^2 + p_3 (x - x_0)^3 + p_4 (x - x_0)^4.
\end{equation}
To constrain this polynomial to be smooth and without inflection points embedded, we now require that the second derivative
\begin{equation}
\frac{\text{d}^2f(x)}{\text{d}x^2} = (2!/0!) p_2 + (3!/1!)p_3 (x - x_0) + (4!/2!)p_4(x-x_0)^2
\end{equation}
has no zero crossings.  Additionally, we also require that the third derivative 
\begin{equation}
\frac{\text{d}^3f(x)}{\text{d}x^3} = (3!/0!) p_3 + (4!/1!)p_4 (x - x_0) 
\end{equation}
has no zero crossings.

In general, we may model the foreground by an $n^{\rm th}$-order polynomial and require that all derivatives of order 2 and higher have no zero crossings within the domain of interest.  This is implemented by computing the functions
\begin{multline}
\frac{\text{d}^mf(x)}{\text{d}x^m} = \{m!/0!\} p_m + \{(m+1)!/1!\} p_{m+1} (x - x_0) + \{(m+2)!/2!\} p_{m+2} (x - x_0)^2 + \\
 \{(m+3)!/3!\} p_{m+3} (x - x_0)^3 +\ .....\ + \{ n! / (n-m)! \} p_n (x - x_0)^{n-m}
\end{multline}
or
\begin{equation}
\frac{\text{d}^mf(x)}{\text{d}x^m} = \sum_{i=0}^{n-m} \{(m+i)! / i! \} p_{m+i} (x - x_0)^i 
\end{equation}
for all $m$ in the range 2, 3, 4, ...., $(n-1)$ and constraining the polynomial coefficients $p_j$ so that there are no zero crossings within the domain for any of these functions.  We call the polynomial functions that satisfies these constraints {\it Maximally Smooth} functions.  They will not have ripples embedded.

We model the foreground in log-temperature versus log-frequency space using a Taylor series expansion about the lowest frequency log$_{10}(\nu_0)$ in our band. The  polynomial is written in terms of powers of $\log_{10}(\nu/\nu_0)$.  If $y(x)$ is the polynomial describing the foreground in log-space, then $10^{y(x)}$ is added to a term that describes the CMB spectrum and another term that models the recombination line spectrum to get a model for the total spectrum. 

We model the recombination line spectrum component in the mock data using a single scaling parameter: the recombination line component is considered to be this scale factor times a template of the spectral ripple that is nominally expected to be present in the observation.  First, the scale factor is in itself a quantitative measure of the recombination template present in the total spectrum.   A scale factor close to unity indicates that the predicted template is present in the spectrum where as a small value indicates the absence of such a spectral signature. The distribution allowed for this scaling parameter by the goodness of fit yields the confidence in the detection of the predicted recombination template.  Secondly, by modeling the recombination lines using a scaled template allows the residuals of the optimization to approach measurement noise if the theoretical framework that led to the template is correct. 

The analytic fitting function in its final form, which includes a Maximally Smooth form for the foreground modeling, is given by 
\begin{equation}\label{eq:fit_func}
T(\nu) = \left( {\frac{h\nu}{k}} \right) / \left( {e^{\frac{h\nu}{kp_{0}}}-1} \right) + p_1T_{\rm rec}(\nu) 
+ 10^{ \sum_{i=0}^{n} \left[\log_{10}(\nu/\nu_0)\right]^i \,p_{i+2}} ,
\end{equation}
where $p_0$ corresponds to the CMB temperature, $p_1$ is the scale factor that multiplies the recombination line template $T_{\rm rec}(\nu)$ and $p_2$ through $p_{2+n}$ are the coefficients of the terms in the $n^{\rm th}$-order Maximally Smooth polynomial that models the foreground. Fig.~\ref{fig:template_ideal} shows the predicted recombination line signature that is expected to be present in the synthetic spectrum as an additive component; as stated earlier, this template has been derived from the predicted recombination line spectrum by subtracting a baseline of low order and is what would be expected as a residual if a smooth baseline were to be subtracted from an observation as a method of removing the foreground. Segments of this template are what form the recombination line template $T_{\rm rec}(\nu)$.


We use the downhill simplex \citep{Nelder1965} optimization algorithm to iteratively fit this model to the synthetic spectrum adopting a successive approximation strategy. We start by fitting to a function that has four coefficients, one for the CMB temperature, one for the amplitude of the recombination line spectrum and two that describe a first order model for the foreground.  We successively include more terms one by one giving an initial guess of zero for each new coefficient.

Adopting the method discussed in Section~3, we have generated synthetic spectra over three octave bands --- 2.0--4.0, 2.5--5.0 and 3.0--6.0~GHz --- spaced uniformly over 24 hr in LST to test the robustness of this modeling algorithm.  In this test, the noise in the spectrum was kept small so that the residual spectrum might reveal the recombination line structure obviously, if recovered successfully.  As expected,  the final fit parameters vary for different realizations of the sky spectrum corresponding to different observation times.   Nevertheless, the algorithm did indeed converge for all LSTs and yielded a best fit scaling factor close to unity at all times.  An 8$^{\rm th}$ order Maximally Smooth polynomial was used to model the foreground.  The residuals were consistent with the measurement noise that had been added to the synthetic spectra and the foregrounds were indeed successfully modeled as 8$^{\rm th}$-order smooth polynomials, without also fitting to the embedded recombination line spectrum.

\begin{figure}[ht]
\centering
\includegraphics[scale = 0.5]{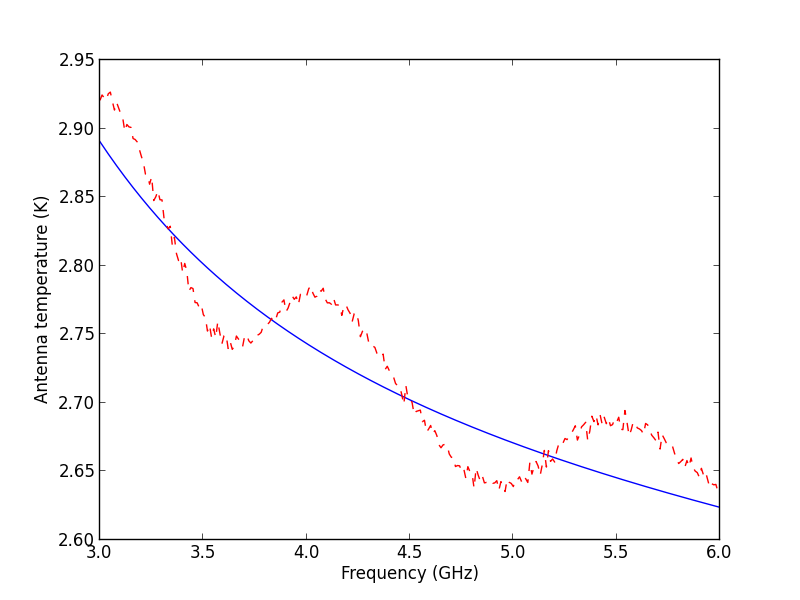} 
\caption{ Foreground model fit to a synthetic spectrum that represents a mock observation is shown in blue. The residual, scaled by a fraction of the inverse of the chi-square of fit and and added back to the foreground model is shown in red. }
\label{fig:foreground_plus_lines}
\end{figure}

\begin{figure}[ht]
\centering
\includegraphics[trim= 20mm 10mm 10mm 10mm, clip,scale = 0.4]{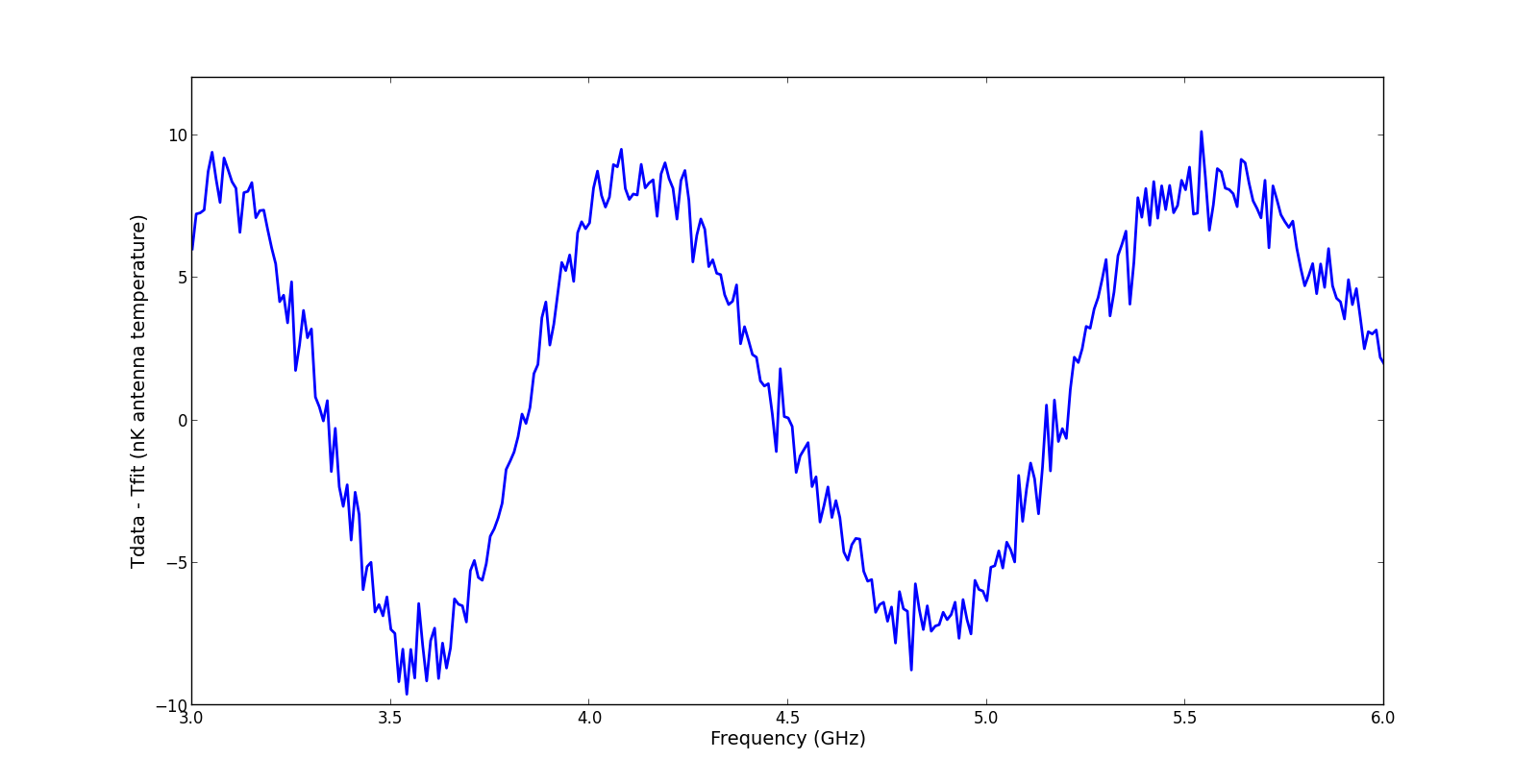} 
\caption{The residual on subtracting the foreground plus CMB model-fit from the synthetic spectrum.}
\label{fig:residual_constrained}
\end{figure}

In Fig.~\ref{fig:foreground_plus_lines} we show a sample fit to a mock observation in the 3--6~GHz band.  The model-fit CMB plus Maximally Smooth foreground is shown along with a second trace that is constructed by computing the residuals to this fit and adding the residuals to the model with a large rescaling.  The plot shows that the model does indeed not fit to the recombination line ripple, which is present in the synthetic spectrum, and does leave the ripple as a residual.  The model fitting algorithm based on the Maximally Smooth polynomials we have constructed is capable of separation of the foreground from embedded ripple, which is nine orders of magnitude smaller. The residual alone is shown in Fig.~\ref{fig:residual_constrained}, which can be compared with Fig.~\ref{fig:unconstrained_residual} where an  8$^{\rm th}$-order polynomial was used without any constraints to fit to the synthetic spectrum.

\begin{figure}[h!]
\begin{centering}
		\begin{minipage}[b]{0.32\linewidth}
			\includegraphics[trim= 5mm 5mm 15mm 0mm,clip,width=\linewidth]{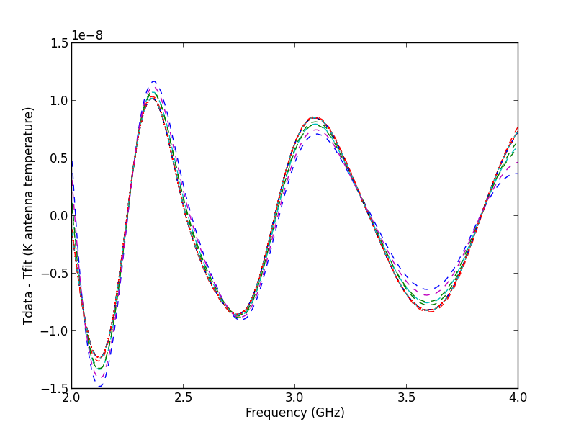}
			\label{fig:2_to_4}
		\end{minipage}
		\begin{minipage}[b]{0.32\linewidth}
			\includegraphics[trim= 5mm 5mm 15mm 0mm,clip,width=\linewidth]{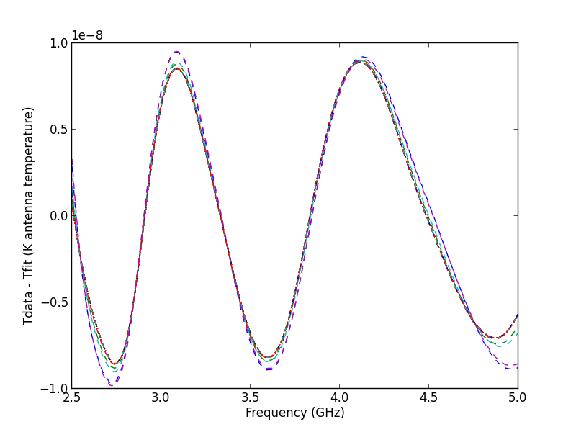}
			\label{fig:2pt5_to_5}
		\end{minipage}
		\begin{minipage}[b]{0.32\linewidth}
			\includegraphics[trim= 5mm 5mm 15mm 0mm,clip,width=\linewidth]{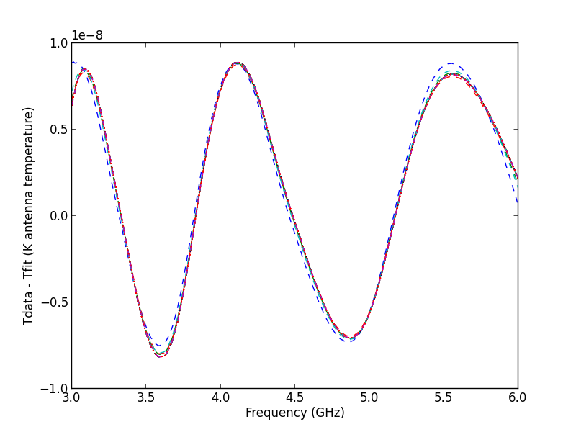}
			\label{fig:3_to_6}
		\end{minipage}
\caption{Residuals in the 2.0--4.0, 2.5--5.0 and 3--6~GHz bands following fitting and subtracting Maximally Smooth foreground models to synthetic sky spectra.  The mock observations are over a 24 hour period, with spectra spaced 2 hours apart, and represent observations made by an ideal instrument as discussed in the text.  Shown as a red dotted line is the recombination line template from the theoretical predictions;  the overlaid colored dashed lines represent the recovered residuals from different realisations of synthetic sky spectra.}
\label{fig:residual_lst}
\end{centering}
\end{figure}

In Fig.~\ref{fig:residual_lst} we show the residuals to the fit for a smooth foreground plus CMB to synthetic spectra for mock observations in different frequency ranges and distributed over LST.   The algorithm does recover the recombination ripple in all cases, giving confidence in the robustness of the method.

 These residuals represent a detection of the cosmological recombination lines.  The recovery demonstrates that the smooth functional form proposed here does not also fit to the nanoscale embedded ripple representing the recombination line signature, despite the functional form being allowed to be of arbitrarily high order.  The recovery demonstrates that it is indeed possible to model fit a smooth functional form--- e.g., the Maximally Smooth functional---to observations and model the completely monotonic foreground---of unknown functional form---to an accuracy better than 10~nK!  

\section{Confidence in Detection}

In the previous section, we presented an algorithm by which we recover cosmological recombination lines of amplitude $\sim$10~nK that are a tiny additive part of a sky spectrum which is nine orders of magnitude brighter. In this section, we first examine the probability distribution for the detected amplitude of the spectral ripple.  We then examine the confidence in detection of the expected ripple and the confidence with which we may reject false positives, using a Bayes factor approach.   These lead to inference of the observing time needed for detections with different confidence levels. 

\subsection{Detection based on the fit amplitude of the recombination spectral ripple}
\label{sec:mcmc}

An approach to determining whether or not a given sky spectrum contains cosmological Hydrogen recombination lines is to jointly fit to the spectrum a model composed of a Planck spectral component for the CMB, a Maximally Smooth polynomial accounting for radiation from point and extended sources in the foreground, and the recombination line template scaled by a factor, which is a variable for the fitting. The functional form for such a model is given by Equation~\ref{eq:fit_func}.
We expect that the fit value of $p_1$, which represents the normalized amplitude of the recombination line ripple, would take on a value of unity if the theory is valid, and a value zero if no lines exist.  We choose 0.5 as the threshold between a null detection and a positive detection, with values smaller than 0.5 assumed to indicate a null detection and greater than 0.5 deemed to be indicating a positive detection. 

To determine the confidence in detection using such an approach, as well as the probability of a false positive, we synthesize two types of mock observations, one in which the theoretically expected cosmological recombination lines are added and another without. Forty spectra of each type were generated with independent thermal noise. Below we adopt the notation that data and associated terms for spectra with recombination lines present in them would be referred to as data set `$a$' and those without recombination lines as `$b$'.  We fit each of the synthetic sky spectra separately with the mathematical model given by Equation~\ref{eq:fit_func} using the successive approximation approach, optimizing all the parameters using the \cite{Nelder1965} algorithm to minimize the Chi-squared difference between the model and mock data, each time increasing the degree of the Maximally Smooth polynomial in Equation~\ref{eq:fit_func} by unity until the root mean square of the residual saturates.

To sample the distribution in scaling factor $p_1$ and thus derive the confidence for detection using a threshold of 0.5 for the scaling factor, we adopt a Markov Chain Monte Carlo (MCMC) analysis.  Specifically, we adopt the EMCEE package \citep{Foreman-Mackey2013}, which is a Python-scripting-language implementation of an affine invariant ensemble sampler as proposed by \cite{Goodman2010}. The {\tt emcee} package wins over traditional implementation of MCMC samplers in its high computational efficiency in generating statistically independent samples from the posterior probability distribution function (PDF). Another major advantage of {\tt emcee} is that it requires tuning of only a couple of parameters by hand in an $N$-dimensional space as opposed to $N^2$ values in other traditional methods. While data driven algorithms do exist to arrive at optimum guesses for initial parameters, for conventional methods this comes at the computational cost of lengthy burn-in chains. 

Chi-squared minimization of the model $T(\nu)$ leads to optimum values for the parameters $p_0$ representing the CMB temperature, $p_1$ representing the scaling factor, and the polynomial coefficients $p_2$, $p_3$, $p_4$, $p_5$ and $p_6$ (for a $4^{th}$ order Maximally Smooth polynomial).   We then invoke the EMCEE sampler with 30 walkers each of which generates a Markov sampling chain for every parameter.  The different chains are initialized at random locations close to the optimum location in the $N=7$ dimensional parameter space.
The EMCEE ensemble sampler operates on the log posterior probability distribution function that we define to be:
\begin{equation}
\ln ({\rm Probability}) = \frac{1}{2}\mathlarger{\mathlarger{\sum}}\bigg(\frac{(T(\nu) - y_{a|b})^2}{\sigma^2} + {\rm ln}(2\pi\sigma^2)\bigg),
\label{eq:lnprob} 
\end{equation}
where $y_{a|b}$ is the synthetic sky spectrum with or without recombination lines. 
We first run a short burn-in of 1000 steps, discard these values and run a second burn-in of 6000 steps.  We ensure that length of the second burn-in is at least ten times the correlation length, then discard these as well and run a final useful 500 steps.  With 30 walkers exploring 500 steps each we generate a distribution of 15000 samples per parameter per synthetic sky spectrum. Marginalizing over the nuisance parameters, which in our case are all parameters with the exception of the scaling factor, we have for 40 sky spectra a distribution of 600,000 scaling factors in all. On running the EMCEE ensemble sampler on data sets `$a$' and `$b$' as described above we have two distributions of scaling factors with 600,000 samples each. 

\begin{figure}[ht]
\centering
\includegraphics[trim=15mm 0mm 20mm 0mm,clip,scale = 0.45]{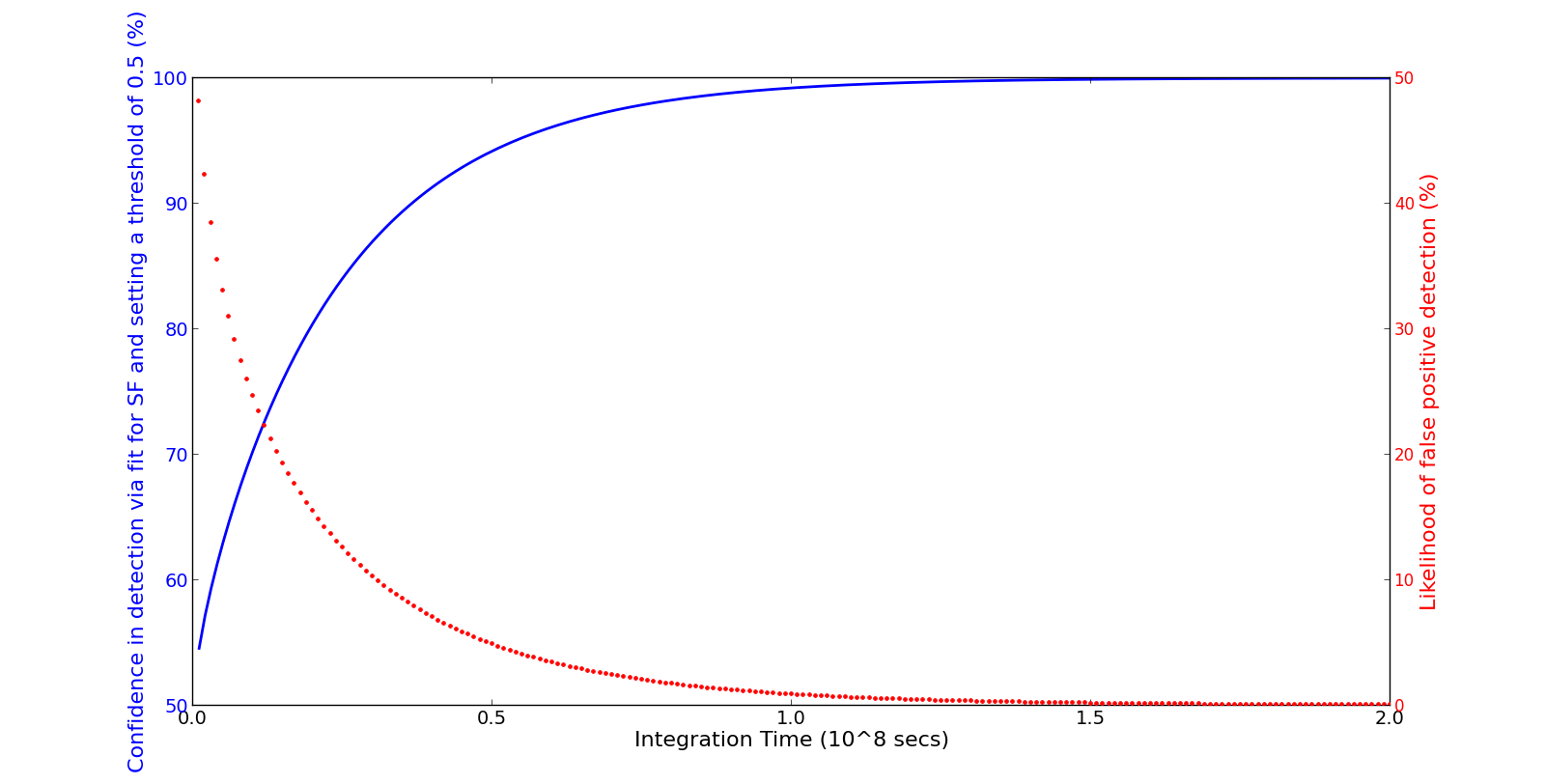} 
\caption{The confidence with which scaling factor (SF) above threshold 0.5 signifies a detection of cosmological recombination lines (shown as a blue solid line with scale on the left of the plot), and percentage likelihood of false positives appearing as a detection for this threshold (shown as a red dotted line with scale on the right side of the plot).  The confidence values and percentage likelihoods of false positives are shown versus integration time assuming cryogenic cooled state-of-the-art 1~K receivers and a 128-element array of precision total-power spectrometers.}
\label{fig:cl_from_sf}
\end{figure}

We generate multiple pairs of data sets  `$a$' and `$b$' with different noise integration times; as a guide we have adopted times corresponding to different signal-to-noise ratios as defined by Equation~\ref{eq:snr}.  For each integration time, the fraction of samples of the scaling factor from the Markov chain corresponding to data set `$a$' that have values exceeding the threshold of 0.5 gives an estimate of the probability of detection.  Similarly the fraction of samples corresponding to data set `$b$' below 0.5 is an estimate of the probability of rejecting a false positive.  These are depicted in Fig.~\ref{fig:cl_from_sf}.  


Detection of the recombination ripple with 68\% confidence using uncooled receivers requires observing with about $210 \times 10^{8}$~antenna secs, or equivalently about $240000$ antenna days. An array of 128 singly-polarized total-power spectrometer elements with uncooled receivers would require about $\sim 5$ years for detection with this confidence and based on $SF$ threshold. However, using cryogenically cooled state-of-the art receivers, the observing time for detection with 68\% confidence is $\sim10 \times 10^{8}$~ antenna secs or, equivalently, 92 observing days with an array of 128 spectral radiometers deploying such cooled receivers. A 90\% confidence detection requires observing with 430 days with such an array, and  95\% confidence detection requires increasing the observing time for such an array to about 625 days.

\subsection{Detection based on Bayes factor (BF)}
\label{sec:bayes_factor}


Consider some data D and a choice between model M$_1$ characterised by a set of parameters  $\theta_1$ and M$_2$ characterised by a set of parameters $\theta_2$, to describe the data. The Bayes factor is given by the ratio of the probability of the data D given the model M$_1$ to the probability of D given M$_2$:
\begin{equation}\label{eq:BF_def}
BF = \frac{P(D|M_1)}{P(D|M_2)} = \frac{\int P(\theta_1|M_1) P(D|\theta_1,M_1) d\theta_1}{\int P(\theta_2|M_2) P(D|\theta_2,M_2) d\theta_2}.
\end{equation}
An alternative expression of the Bayes factor is in terms of posterior and prior odds. Rewriting the probabilities associated with the two models as odds, we obtain:
\begin{equation}\label{eq:BF_def2}
  \begin{split}
    {\rm Posterior\ odds} = BF \times {\rm Prior\ odds}, \\
    {\rm or\ that:~~}\frac{P(M_1|D)}{P(M_2|D)} = BF \times \frac{P(M_1)}{P(M_2)}, \\
    {\rm which\ leads\ to:~~}BF = \frac{P(M_1|D)}{P(M_2|D)}\times \frac{P(M_2)}{P(M_1)}.
  \end{split}
\end{equation}
When the prior odds are 1:1 the Bayes factor reduces to the likelihood ratio.

We once again adopt the notation used in Section~\ref{sec:mcmc} so that symbols representing data and associated terms for mock observations with the recombination lines present in them would be given subscript `$_a$' and those without recombination lines `$_b$'.  Terms representing the null hypothesis would have subscript `$_0$' and alternative hypothesis `$_2$'.  In our problem of estimating the confidence in the detection of recombination line ripples in observed spectra, we consider two hypotheses. The first hypothesis is that the spectrum does not contain recombination lines of any detectable amplitude; {\it i.e.}, within errors the mock spectrum can be adequately modeled by the total of contributions from the cosmic microwave background and a Maximally Smooth function representing the cumulative spectra from diffuse and point sources in the foreground. We refer to this as the null hypothesis $H_0$. The second hypothesis, which we refer to as the alternative hypothesis $H_2$, is that the observed sky spectrum includes the cosmological Hydrogen recombination lines as an additive component along with contributions from the CMB and foregrounds.  Hypothesis $H_2$ is that the spectrum in question does contain the recombination lines exactly as given by the theoretical expectations.  The Bayes factor for a comparison between these alternate hypotheses is given by:
\begin{equation}\label{eq:BF}
BF = \frac{P(D|H_2)}{P(D|H_0)}\times\frac{P(H_2)}{P(H_0)},
\end{equation}
where $P(D|H_2)$, $P(D|H_0)$ are the respective likelihoods of the alternative and null hypotheses and $P(H_2)$, $P(H_0)$ are the prior probabilities of the alternative and null hypotheses. Since in this comparison we have no {\it a priori} preference or bias towards the presence or absence of the recombination ripple in data, we may assume equal prior probabilities and hence the Bayes factor reduces to the likelihood ratio given by:
\begin{equation}\label{eq:BF_2}
	BF = \frac{P(D|H_2)}{P(D|H_0)}.
\end{equation}

We now proceed to compute the likelihood functions for the above two hypotheses.  We generate data set `a', a set of 100 mock observations with independent thermal noise and with each containing the recombination lines as an additive component. We then fit each of these spectra independently with two different models. The first model is the null hypothesis $H_0$ with a blackbody function to model the CMB component and a Maximally Smooth polynomial to model the foreground component of the sky spectrum. $H_0$ expects that these two components completely describe the spectrum and that the residual on subtracting this model-fit from the data must be Gaussian random with zero mean and standard deviation given by the measurement noise. We compute the likelihood of obtaining the data given $H_0$ by: 
\begin{equation}\label{eq:likeli_1}
P(D_a|H_0) = \prod_{i=1}^{N}\frac{e^{-\frac{y_{res0}[i]^2}{2\sigma_0^2}}}{\sqrt{2\pi\sigma_0^2}},
\end{equation}
where $N$ is the number of independent points across the spectrum and $y_{res0}[i]$ is the residual spectrum following subtraction of the model corresponding to the null hypothesis.  

The variance of the measurement noise, $\sigma_0^2$, is estimated from the data itself.  This variance is assumed to be half the Allan variance (AV) of the residual, which is given by:
\begin{equation}\label{eq:av}
AV = \sum_{i=1}^{N-1}(y_{res}[i+1]-y_{res}[i])^2,
\end{equation}
where $y_{res}$ is the residual on subtracting the model from the data.  Since in our  cases the signal-to-noise is small in the channel data, this approach makes the estimate for measurement error robust and independent of errors in the model fit and any low-order residuals that are not represented in the model.  

\begin{figure}[ht]
\centering
\includegraphics[trim = 0mm 0mm 10mm 5mm,clip,scale = 0.5]{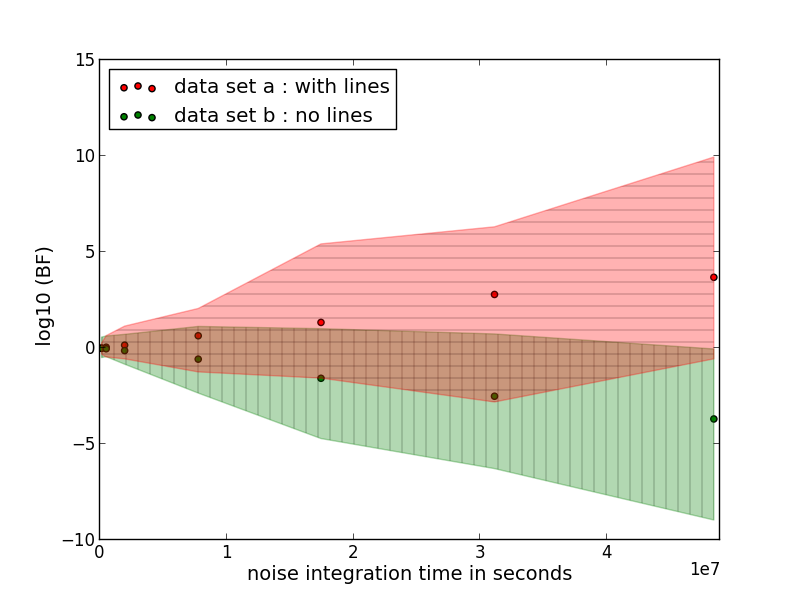} 
\caption{Bayes factors computed for data sets `a' and `b' that are mock observations generated with and without the recombination line component  respectively.   log$_{10}(BF)$ is plotted versus observing time assuming cryogenic cooled state-of-the-art 1~K receiver noise temperatures and a 128-element array of precision total-power spectrometers. Filled circles represent the median values of the $BF$'s computed separately for the two data sets. The horizontally-striped region shaded in red represents the range in Bayes factors obtained for the 100 mock observations in data set `a' and the vertically-striped region shaded in green represents the range of Bayes factors obtained for data set `b'.}
\label{fig:bf_time}
\end{figure}

\begin{figure}[ht]
\centering
\includegraphics[trim = 5mm 5mm 10mm 5mm,clip,scale = 0.55]{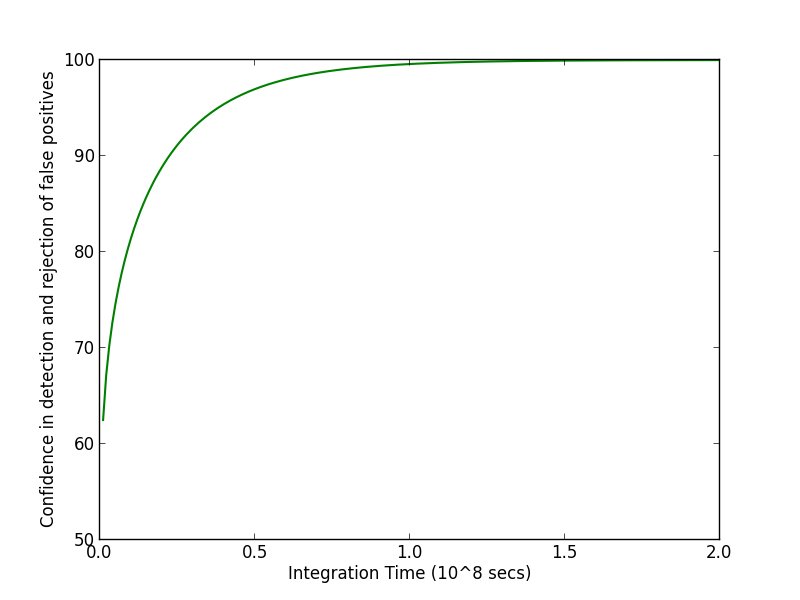} 
\caption{The confidence with which Bayes factor above unity might signify a detection, and simultaneously reject the detection to be a false positive, versus integration time. We assume cryogenic-cooled state-of-the-art 1~K receiver noise temperatures and a 128-element array of precision total-power spectrometers.}
\label{fig:mean_cl}
\end{figure}

The second model corresponds to the alternative hypothesis $H_2$.  This model contains the blackbody CMB term, the Maximally Smooth polynomial representing the foreground and, in addition, a template of the recombination lines that are expected to be present in the spectra for this hypothesis. As before, we proceed by fitting the data with the model, subtracting the best fit model from the data and computing the likelihood from the residual. We again make the reasonable assumption that if the data is completely represented by the model then the residual should be Gaussian random noise corresponding to measurement noise. The likelihood function P($D_a|H_2$) is given by:
\begin{equation}\label{eq:likeli_2}
P(D_a|H_2) = \prod_{i=1}^{N}\frac{e^{-\frac{y_{res2}[i]^2}{2\sigma_2^2}}}{\sqrt{2\pi\sigma_2^2}},
\end{equation}
where $y_{res2}[i]$ is the residual on subtracting the alternative hypothesis model from the data, and variance $\sigma_2^2$ is half the Allan variance of this residual.  Since the Allan variance gives the measurement noise independent of the model used to fit to the data, we expect $\sigma_0^2$ and $\sigma_2^2$ to be the same.

Having computed the likelihoods for both the models we arrive at the Bayes factor by computing the likelihood ratio:
\begin{equation}\label{eq:BF_w_line}
BF_a = \frac{P(D_a|H_2)}{P(D_a|H_0)}.
\end{equation}

We repeat this exercise with a separate data set `b', which is a set of 100 spectra corresponding to mock observations with the same measurement noise variance as in data set `a', but with no recombination lines in the synthetic spectra. We once again fit these spectra with the two models as before. We compute the likelihood functions for the null and alternative hypotheses given by $P(D_b|H_0$) and $P(D_b|H_2$) respectively, under the reasonable assumption that when a model that completely represents the data, except for random noise, is subtracted from the data the residual must be the measurement noise.  Thus the Bayes factor for the 100 mock observations that are generated assuming that recombination line component is absent is given by:
\begin{equation}\label{eq:BF_n_line}
BF_b = \frac{P(D_b|H_2)}{P(D_b|H_0)}.
\end{equation}

Fig.~\ref{fig:bf_time} shows a plot of log$_{10}$($BF_a$) and log$_{10}$($BF_b$) versus integration time.  With increasing integration time the measurement noise reduces and the two Bayes factors diverge, as expected.

For any data set, the Bayes factor indicates the relative preference of the data for the two hypotheses.  The median $BF_a$ rises above unity with increasing integration time, increasingly preferring the hypothesis $H_2$ over $H_0$.  In contrast the median $BF_b$ drops below unity with increasing integration time, increasingly preferring the hypothesis that the recombination lines are absent in data set `b'.  We may set a threshold at unity and examine the confidence with which this threshold might discriminate between observations that contain a recombination ripple to those in which this additive component is absent.  At any integration time, the distribution of Bayes factors indicates the probability that $BF_a$ is above unity, which gives the confidence in a detection.  For the same integration time, the distribution of $BF_b$ yields the probability of obtaining a value above unity, which gives the likelihood of a false positive.  Using the multiple data sets `a' and `b' we have computed the run of the confidence in detection (and simultaneous rejection of false positives) versus integration time: this is depicted in Fig.~\ref{fig:mean_cl}.


If we adopt the Bayes factor as a detection statistic, detection of the recombination ripple with 68\% confidence using uncooled receivers requires observing with about $52 \times 10^{8}$~antenna secs, or equivalently about $60,000$ antenna days. An array of 128 singly-polarized total-power spectrometer elements with uncooled receivers would require about $1.3$ years for detection with this confidence. Using cryogenically cooled state-of-the art receivers, the observing time for such detection with 68\% confidence is $2.56 \times 10^{8}$~ antenna secs or, equivalently, $23$ observing days with an array of 128 spectral radiometers deploying such cooled receivers.  90\% confidence detection requires observing with 255 days with such an array, and  95\% confidence detection requires about 440 days.

\begin{figure}[ht]
\centering
\includegraphics[trim = 15mm 5mm 10mm 5mm,clip,scale = 0.5]{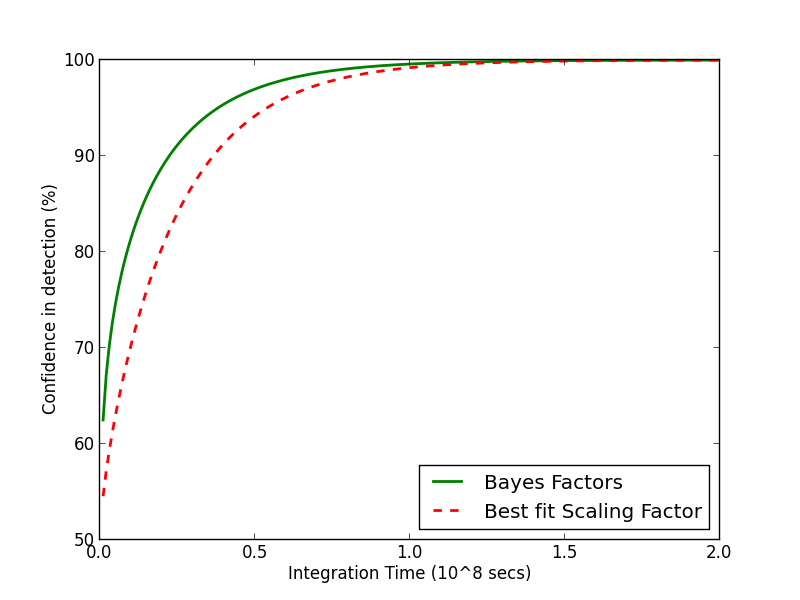} 
\caption{A comparison of the confidence in detecting cosmological recombination lines in sky spectra using Bayes Factors and that given by a comparison of the best-fit ripple amplitude to a threshold. The confidence values are shown versus integration time assuming cryogenic cooled state-of-the-art 1~K receiver noise temperatures and a 128-element array of precision total-power spectrometers.}
\label{fig:mcmc_vs_bf}
\end{figure}

Fig.~\ref{fig:mcmc_vs_bf} shows the confidence in the detection of cosmological recombination lines in synthetic sky spectra as a function of integration time, estimated using both the Bayes factor method as well as using the fit to the amplitude of the recombination ripple (Section~\ref{sec:mcmc}). The Bayes factor approach to detection is simplest in that it only asks whether in an observed spectrum the lines as predicted by theory are present or totally absent. The method that derives an amplitude for the ripple, as estimated by a best fit to a scaling factor for a template of the expected ripple, attempts to ask a more informative question in that it attempts to evaluate the amplitude of the recombination line ripple. Unsurprisingly, the Bayes factor method has greater confidence in its answer for any integration time.

\section{Summary}

We demonstrated that it is in principle feasible with current day technology to experimentally detect the cosmological Hydrogen and Helium recombination lines at low frequencies, although the lines are embedded in a foreground that is about nine orders of magnitude brighter with a priori unknown precise spectral shape.  The recombination radiation has a smooth component that is difficult to distinguish from the smooth foregrounds; however, the recombination radiation also has a unique ripply component that may be distinguished from the foregrounds and instrumental effect. We estimate the amplitude of the ripply signal, the instrument noise arising from receivers and foregrounds, and estimate the signal-to-noise in detections using ground based spectrometers.  Detection may ideally be attempted using an octave band in the 2--6~GHz window; an octave band would have a spectral segment of the recombination ripple with sufficient structural complexity so as to be distinguishable from the relatively smoother foreground.  We have developed an algorithm to detect the recombination line ripple by foreground modeling that models the foreground as a {\it Maximally Smooth} polynomial, which we define enforcing no zero crossings in derivatives of order $n\geq 2$. A similar approach may be applicable to modeling and detecting the global 21-cm signal at lower frequencies.

We then evaluate the confidence with which these cosmological recombination lines may be isolated using the aforementioned algorithm and estimate the integration times required for detection with varying degrees of confidence. In its simplest form, a detection is testing an observed spectral segment addressing the question of whether or not the theoretically motivated spectral ripple is present in the sky spectrum or absent. To answer this question with 90\% confidence requires an integration time of  32,640 antenna-days with a total-power spectral radiometer with cryogenically cooled state-of-the art receivers; however, with today's best uncooled receivers the integration time for such a detection increases to as much as 660,000 antenna days.  This makes a compelling case for the use of cryogenically cooled receivers.  Moreover, since the antenna element for such all-sky global signals does not gain from antenna directivity,  very small antenna elements of centimeter dimensions may be deployed, perhaps as a compact cluster of antenna elements housed in a moderate-size dewar.  Assuming a 128 element array followed by independent spectral radiometers, a 90\% confidence detection may be achieved in 255 days integration time.

APSERa\footnote{http://www.rri.res.in/apsera/}---Array of Precision Spectrometers for the Epoch of RecombinAtion---is an experimental venture at the Raman Research Institute\footnote{http://www.rri.res.in/}, India, with the science goal of detecting spectral ripples from the epochs of Hydrogen and Helium recombination. On completion, it is to be an array of 128 miniature radio telescopes operating in the octave band of 3--6~GHz with antenna elements and receivers that are custom designed for all-sky spectral measurements with the required sensitivity to experimentally detect these cosmological recombination lines. APSERa will be deployed at a radio quiet site closer to the geographic poles to avoid radio frequency interference in these bands from downlinks of geostationary satellites.

\section{Future Investigations}

Ripple like spectral signatures arising from the epochs of Hydrogen and Helium recombination are predicted to appear as additive distortions to the sky spectrum at a level $\sim$9 orders of magnitude weaker than the total brightness of the radio sky. We have demonstrated that it is in principle possible to detect these signals. Due to their small amplitude, any small departure in the spectral radiometer from the ideal behavior assumed herein could critically affect the detection likelihood. Studies of the effects of potential instrumental non-idealities, and confusion arising from additive astrophysical contaminants, which have not been considered here, will form the next step in our future work. The key investigations are listed below.
\begin{itemize}
\item The spectral radiometer radio telescope used to detect the weak cosmological recombination lines must be capable of achieving the required spurious-free dynamic range so as to recover the embedded cosmological signal in the measurement.  This implies that the receiver configuration and calibration techniques must be specifically designed to minimize systematics and that any residual systematics are either well characterized or emendable to accurate modeling so as to enable unambiguous detection of the weak signal.  A future work is the design of a suitable receiver configuration, observing strategies, gain stability and calibration methods to achieve the required accuracy in the measurement. 
\item In our simulations, we have generated mock sky spectra as observed by an ideal frequency independent antenna beam.  The spectral response of such an antenna is {\it maximally smooth} if all sources in the beam have spectra that are complete monotones.  However, designing and fabricating an antenna that is frequency independent over an octave bandwidth is non-trivial.  Frequency dependence in the beam pattern and its sidelobes may result in a response that is no longer smooth and may confuse the detection of cosmological recombination.  A study of this mode coupling of sky structure into spectral structure in a spectral radiometer, which depends on the type of frequency dependence in the beam pattern, would lead to design tolerances on the antenna element for this detection experiment.   
\item The all-sky `ripply' signal arising from the epochs of Hydrogen and Helium recombination is inherently unpolarized and this spectral feature may be detected with an antenna that responds to any single polarization mode, either linear or circular.  However, the foreground synchrotron emission from extragalactic sources as well as Galactic emission is linearly polarized and Faraday rotation during the line-of-sight propagation results in the received polarization position angle varying with observing frequency.  This causes linearly polarized sources to appear with a `ripply' spectral structure in the response of linearly polarized antennas.  One possible way of eliminating beam asymmetries and the effect of beam rotation across the sky is to rotate the array itself. A study of this mode coupling of polarized sky emission to spectral structure is a future study that will lead to design tolerances on the polarization properties of the antenna element and on the polarization calibration. The potential of using the unpolarised nature of the cosmological signal to discriminate it from polarised foregrounds is another aspect to be explored.
\item The spectral template of the additive ripple-like feature from cosmological Hydrogen and Helium recombination has a fairly accurate theoretical prediction. The near quasi-sinusoidal nature of this signal acts as a fingerprint providing a means to distinguish it from other weak cosmological signals. Two such cosmological signals are the $\mu$ and $y$ distortions of the CMB \citep{Zeldovich1969, Sunyaev1970mu}. Even within the standard cosmological model, these distortions are created at amplitudes that are typically larger than the cosmological recombination radiation \citep[e.g.,][]{Chluba2011therm, Sunyaev2013, Chluba2013fore, PRISM2013WPII}. A large average Compton $y$ distortions, at the level of $y\simeq 10^{-7}-10^{-6}$, is expected from reionization and structure formation \citep{Sunyaev1972b, Hu1994pert, Cen1999, Oh2003} and unresolved clusters and filaments \citep{Refregier2000, Miniati2000, Zhang2004}.  At low frequencies, this type of distortion only leads to a weak frequency dependent signal, $T(\nu)\simeq -2 y (1-x^2/12)$ with $x\approx 0.018 \, [\nu/\GHz]$, making it is less problematic. For a $\mu$ distortion, the low frequency spectrum shows much richer structure, especially when varying time-dependence of the energy release mechanism \citep[Fig.~14,][]{Chluba2011therm}. Each of these introduce spectral features in the CMB with the $\mu$ distortion peaking at about 1~GHz. In an experiment optimized to detect ripples originating from the epochs of Hydrogen and Helium recombination it would be interesting to study whether $\mu$ and $y$ distortions of the CMB are a possible source of confusion or would themselves be detected as a positive by-product. Similarly, the anomalous microwave emission from spinning dust \citep[see][]{Draine1998,Haimoud2009,Planck2014} will add another spectral dependences to the low-frequency regime, which at this point we have not investigated.
\end{itemize}

\section{Acknowledgements}
The authors would like to thank Franklin H. Briggs at the Research School for Astronomy \& Astrophysics, Australian National University, for his valuable inputs throughout the preparation of this paper. JC is supported by NSF Grant No. 0244990. We acknowledge the ARC Centre of Excellence for All-sky Astrophysics (CAASTRO) and the Australian National University for support under the distinguished visitor program. 

\appendix
\section{The Completely Monotone (CM) function}\label{sec:appendix_a}
 {
Mathematically, a function $f(x)$ is said to be a complete monotone (CM) if for all values of $x$ in the interval $0 \le x < \infty$,  $(-1)^n\times\frac{\textrm{d}^nf(x)}{\textrm{d}x^n} \ge 0$ for every integer $n \ge 0$. The method by which a CM function of any said order may be constructed is best explained by an example, such as that for a cubic function as given below. We construct a cubic function $f(x)$ which is CM in the range defined by $[x_0, x_m]$:
\begin{equation}
f(x) = p_0 - p_1 (x -x_0) + p_2 (x - x_0)^2 - p_3 (x - x_0)^3,
\end{equation}
where $p_i$ for all $0\le i \le3$ are the coefficients that are constrained 
so that $(-1)^n\times\frac{\text{d}^nf(x)}{\text{d}x^n} \ge 0$ for every integer $n\ge0$ and for all $x$ in $[x_0, x_m]$.
The third derivative of $f(x)$ is 
\begin{equation}
\frac{\text{d}^3f(x)}{\text{d}x^3} = -6p_3.
\end{equation}
To constrain $f(x)$ to be CM we require that $(-1)^3\times\frac{\text{d}^3f(x)}{\text{d}x^3} \ge 0$, or that $p_3 \ge 0$.  Let $p_3 = a_3$, where $a_3$ is a positive value.
The second derivative of $f(x)$ is 
\begin{eqnarray*}
\frac{\text{d}^2f(x)}{\text{d}x^2} & = & 2p_2 - 6p_3 (x - x_0)\\
& \equiv & 2p_2 - 6a_3 (x - x_0)
\end{eqnarray*}
Once again to constrain $f(x)$ to be CM we require that $(-1)^2\times\frac{\text{d}^2f(x)}{\text{d}x^2} \ge 0$, which leads to the condition that $2p_2 - 6a_3 (x - x_0) \ge 0$.  This is satisfied if $p_2$ is selected such that $2p_2 - 6a_3 (x_m - x_0) \ge 0$, which ensures that the criterion is satisfied for all $x$ in $[x_0, x_m]$.  Therefore, we express $p_2$ as $p_2 = a_2 + 3a_3 (x_{max} - x_0)$,  where $a_2 > 0$.
The first derivative of $f(x)$ is 
\begin{eqnarray*}
\frac{\text{d}f(x)}{\text{d}x} & = & - p_1 + 2p_2 (x - x_0) - 3p_3 (x - x_0)^2\\
& \equiv & - p_1 + 2\{a_2 + 3 a_3 (x_m - x_0)\}(x - x_0)  - 3 a_3 (x - x_0)^2
\end{eqnarray*}
To constrain $f(x)$ to be CM, we once again require that $(-1)\times\frac{\text{d}f(x)}{\text{d}x} \ge 0$.  This requires that we determine $p_1$ such that $- p_1 + 2(a_2 + 6a_3 \{x_m - x_0)\} (x_m - x_0)  + 3(a_3) (x_m- x_0)^2 \ge 0$, ensuring that the criteria is satisfied for all $x$ in $[x_0, x_m]$.  To guarantee this we express $p_1$ as $p_1 = a_1 + 2 a_2 (x_m - x_0) + 3 a_3 (x_m - x_0)^2 $, where $a_1 > 0$.
Since the constant term $p_0$ represents a vertical translation of the function and does not affect the shape---the smoothness of the function itself---we allow it to be a free parameter $a_0$.  
In summary, we describe the cubic CM function to have the form 
\begin{equation}
f(x) = a_0 - \{a_1 + 2 a_2 (x_m - x_0) + 3 a_3 (x_m - x_0)^2\} (x -x_0) + \{a_2 + 3 a_3 (x_m - x_0)\} (x - x_0)^2 - a_3 (x - x_0)^3,
\end{equation}
where all of the coefficients $a_0$, $a_1$, $a_2$ and $a_3$ are non-negative.  In this functional form, setting $a_3=0$ reduces the function to CM of order 2 and if we further set $a_2=0$ we have a CM polynomial of order unity.  This formulation allows a successive approximation to data by first fitting for a CM polynomial of order unity as a Taylor expansion at the reference point $x = x_0$ to optimize for coefficients $a_0$ and $a_1$, then adding a higher order term with initial guess $a_2=0$ and optimizing all three coefficients, and finally adding a cubic term with initial guess $a_3=0$ and optimizing all four coefficients.  All coefficients are constrained to be positive by expressing them in the form $a_i = 10^{b_i}$, where the $b_i$'s are allowed to be real numbers.\\\\
Thus a CM polynomial of arbitrary order $n$ may be written in the form
\begin{equation}
f(x) = a_0 + \sum_{i=1}^{n} (-1)^i (x - x_0)^i \{ \sum_{j=0}^{n-i} a_{i+j} C^{i+j}_{j} (x_m - x_0)^{j} \},
\end{equation}
where $C^n_k$ denotes the binomial coefficient $n!/\{k! (n-k)!\}$ and the constant term $a_0$ is left unconstrained so as to allow arbitrary vertical translations. 
}

\end{document}